\documentclass[sigconf]{aamas} 

\usepackage{balance} 
\usepackage{algorithm}
\usepackage{algpseudocode}
\usepackage{graphicx}
\usepackage{bbm}
\usepackage{dsfont}
\usepackage{relsize}
\newcommand{\argmax}{\arg\!\max}

\setcopyright{ifaamas}
\acmConference[AAMAS '23]{Proc.\@ of the 22nd International Conference
on Autonomous Agents and Multiagent Systems (AAMAS 2023)}{May 29 -- June 2, 2023}
{London, United Kingdom}{A.~Ricci, W.~Yeoh, N.~Agmon, B.~An (eds.)}
\copyrightyear{2023}
\acmYear{2023}
\acmDOI{}
\acmPrice{}
\acmISBN{}

\acmSubmissionID{1060}

\title[IQ-Flow]{IQ-Flow: Mechanism Design for Inducing Cooperative \\
           Behavior to Self-Interested Agents in Sequential Social Dilemmas}

\author{Bengisu Guresti}
\affiliation{
  \institution{Istanbul Technical University \\ ITU AI Center}
  \city{Istanbul}
  \country{Turkey}}
\email{guresti15@itu.edu.tr}

\author{Abdullah Vanlioglu}
\affiliation{
  \institution{Istanbul Technical University \\ ITU AI Center}
  \city{Istanbul}
  \country{Turkey}}
\email{vanlioglu16@itu.edu.tr}

\author{Nazim Kemal Ure}
\affiliation{
  \institution{Istanbul Technical University \\ ITU AI Center}
  \city{Istanbul}
  \country{Turkey}}
\email{ure@itu.edu.tr}

\begin{abstract}
Achieving and maintaining cooperation between agents to accomplish a common objective is one of the central goals of Multi-Agent Reinforcement Learning (MARL). Nevertheless in many real-world scenarios, separately trained and specialized agents are deployed into a shared environment, or the environment requires multiple objectives to be achieved by different coexisting parties. These variations among specialties and objectives are likely to cause mixed motives that eventually result in a social dilemma where all the parties are at a loss. In order to resolve this issue, we propose the Incentive Q-Flow (IQ-Flow) algorithm, which modifies the system's reward setup with an incentive regulator agent such that the cooperative policy also corresponds to the self-interested policy for the agents. Unlike the existing methods that learn to incentivize self-interested agents, IQ-Flow does not make any assumptions about agents' policies or learning algorithms, which enables the generalization of the developed framework to a wider array of applications. IQ-Flow performs an offline evaluation of the optimality of the learned policies using the data provided by other agents to determine cooperative and self-interested policies. Next, IQ-Flow uses meta-gradient learning to estimate how policy evaluation changes according to given incentives and modifies the incentive such that the greedy policy for cooperative objective and self-interested objective yield the same actions. We present the operational characteristics of IQ-Flow in Iterated Matrix Games. We demonstrate that IQ-Flow outperforms the state-of-the-art incentive design algorithm in Escape Room and 2-Player Cleanup environments. We further demonstrate that the pretrained IQ-Flow mechanism significantly outperforms the performance of the shared reward setup in the 2-Player Cleanup environment.
\end{abstract}


\newcommand{\BibTeX}{\rm B\kern-.05em{\sc i\kern-.025em b}\kern-.08em\TeX}

\begin{document}




\maketitle 


\section{Introduction}

Social Dilemmas \cite{Rapoport1974GameTA} emerge when self-interested parties have conflicting objectives. Greed or fear of being exploited drives the agents towards defecting, which results in worse outcomes for the whole group in comparison to outcomes that would come out of cooperation \cite{pyschologydilemma, leibo2017multi}. This problem has many applications in computer science, economics and social sciences; hence, it is well-studied under Game Theory using Matrix Game Social Dilemmas (MGSD) and their iterated extension Repeated Matrix Games \cite{leibo2017multi}. Although MGSDs are useful for modelling social dilemmas in real world scenarios, they omit significant characteristics of real world social dilemmas, which are addressed by Sequential Social Dilemmas (SSD) due to their temporally extended structure \cite{leibo2017multi}. Since cooperation and defection are defined for policies in SSD rather than elementary actions \cite{leibo2017multi}, how to induce cooperative behavior to agents in an SSD while the agents are concurrently learning is an open research question. 

Centralized training methods \cite{foerster2018counterfactual, sunehag2017value, rashid2018qmix} are popular approaches in Multi Agent Reinforcement Learning (MARL) when cooperation is necessary. However,  the centralized approaches involve a shared objective to optimize agents' policies and assume full control over agents' internal parameters and learning. Nevertheless, as the use of artificial intelligence becomes common and agents that are separately trained for different objectives are deployed in a shared environment \cite{yang2021adaptive}, it will not be realistic to either expect no conflicting objectives or assume full control over agents' internal parameters and learning. Since it is not possible to guarantee the type, tasks and number of the deployed agents in an unrestricted environment, agents need to be able to continually learn and adapt to the environment while cooperating with each other. Therefore, in this work, we focus on independently learning self-interested agents in an SSD where the agents receive adaptive incentives in order to promote cooperation.

There are different configurations for how agents can be incentivized during learning. Agents can give each other adaptive incentives to shape each other's behavior for their own benefit \cite{foerster2017learning, letcher2018stable}, or there can be a central institution to provide the incentives to shape the agents' behavior for the welfare of the whole community \cite{Baumann2020AdaptiveMD, yang2021adaptive}. In this work, we adopt the latter approach and provide a mechanism that provides incentives to all of the agents in the system in order to prevent any undesirable outcome, such as tragedy of the commons due to defecting. While it might seem like a trivial problem to learn incentives for the mechanism, since providing the average reward to all agents would certainly remove the existing dilemma, it is shown to yield suboptimal results \cite{subanarchy, yang2021adaptive}. Therefore, it is important to design a mechanism that promotes cooperation without incurring performance losses. Furthermore, promoting cooperation to artificial self-interested agents is not the only direction for mechanism design research. Mechanism design can also be used to model human incentives and solve human dilemmas such as determining tax rate for a higher social welfare \cite{yang2021adaptive}. 

In this work, we propose Incentive Q-Flow (IQ-Flow) algorithm to design incentive mechanisms for increasing social welfare and promoting cooperation. IQ-Flow aims to make the cooperative policy correspond to the self-interested policy of the agents by changing system's reward setup. IQ-Flow collects the experience obtained from agents into a replay buffer and trains critic networks to learn state-action values (Q-Values) for agents' self-interests and the group's collective interest. IQ-Flow parameterizes incentive function using meta-parameters and performs meta-gradient learning as in \cite{metagrad, calian2021balancing, LIO, yang2021adaptive} to update the incentive network. In order to learn incentive meta-parameters, IQ-Flow trains the critic using Offline Implicit Q-Learning \cite{implicitqlearning} with the train set for multiple steps and obtains updated parameters, performs policy evaluation with the validation set, and updates the meta-parameters in the direction that makes the actions of the collaborative policy the greedy choice for self-interested agents' Q-Values.

Our algorithm is distinguished from the existing incentive design methods by grounding itself on reward system shaping rather than opponent shaping. Using Offline Reinforcement Learning (Offline RL) with Implicit Q-Learning makes it possible to get a proximate estimate of Q-Values for as greedy as possible policies with self-interested and collective interest objectives only using experience collected by external agents. This approach enables IQ-Flow to modify the reward system with the incentive function by getting a close enough estimate of how changing the incentive affects the expected future return brought about by the reward system. Using an offline method such as Implicit Q-Learning instead of standard Deep Q-Learning is justified by the fact that incentivizer critic has an indirect effect on recipient agents' policies and can only affect collecting experience indirectly. Furthermore, using Implicit Q-Learning also makes extending IQ-Flow to fully offline training simpler for future work. As opposed to opponent shaping based algorithms, IQ-Flow does not possess or make assumptions on any of the agents' internal parameters, learning algorithms, or hyperparameters which makes it independent from the agents in the environment except for the collected experience. Another key difference of IQ-Flow from existing work is that it does not require cost regularization to train an incentive mechanism in SSDs. Nevertheless, we find that including cost regularization improves IQ-Flow's performance as well. Finally, it should be noted that IQ-Flow does not learn a multi-agent policy or perform value factorization to determine the actions of a cooperative policy. This is due to the fact that the algorithm only needs to know the cooperative or selfish action of a specific agent when the actions of other agents are provided. Our contributions can be summarized as below:

\begin{itemize}
    \item Proposing reward system shaping instead of opponent shaping for incentive design; thus, instead of pushing agents towards a Nash-Equilibrium with cooperative outcomes, modifying the reward system such that rational agents are stuck in Nash-Equilibrium with cooperative outcomes
    \item Extending incentive design framework to learn mechanisms off-policy using offline RL and replay buffer; thus, applying offline RL and replay buffer with meta-gradient learning for MARL for the first time to the best of our knowledge
    \item Removing the requirement of accessing or making assumptions on agents' internal learning state for incentive design
    \item Removing the requirement of cost regularization for incentive design in SSDs
\end{itemize}
We illustrate how IQ-Flow operates for Iterated Matrix Games in Iterated Prisoner's Dilemma, Iterated Chicken Game and Iterated Stag Hunt. We further evaluate the performance of our algorithm in the common benchmarks Escape Room \cite{LIO} and SSD-Cleanup \cite{SSDOpenSource, hughes2018inequity, jaques2019social} with 2 Players. We demonstrate that it outperforms the state-of-the-art incentive design algorithm ID and perform ablation studies for IQ-Flow. We further demonstrate that the pretrained mechanism, learned by IQ-Flow, leads to significantly better learning performance than using a shared reward setup. We provide the code for our implementation and experiments at \href{https://github.com/data-and-decision-lab/IQ-Flow.git}{https://github.com/data-and-decision-lab/IQ-Flow.git}.


\section{Related Work}

Centralized training methods in MARL such as COMA \cite{foerster2018counterfactual}, VDN \cite{sunehag2017value}, and QMIX \cite{rashid2018qmix} are successful at optimizing all agents' policies or factorize value functions to achieve a common objective. However, SSD problems can not be approached as fully cooperative problems due to the nature of the problem emerging from coexisting mixed motives and diverse objectives. Hence, decentralized training methods have been developed along with opponent shaping and incentivization practices \cite{foerster2017learning, letcher2018stable, jaques2019social, LIO} in order to model and resolve social dilemma problems.

Opponent shaping was proposed by \cite{foerster2017learning} to provide independent learners with the ability to shape each other's behavior in the face of a mixed motive. LOLA \cite{foerster2017learning} agents can access the policy parameters of their opponents and actively learn in the direction that improves their own returns by considering how their opponent's future policy is expected to change. The disadvantage of the LOLA is that it can adopt arrogant behavior, as claimed by \cite {letcher2018stable} and fixed with a new algorithm named SOS. SOS \cite {letcher2018stable} algorithm is similar to LOLA in adopting opponent shaping, but offers a more robust algorithm by removing the arrogant behavior and inheriting the guarantees of LookAhead \cite{policypred} on avoiding strict saddles in all differentiable games. 

Incentivization practices can be exemplified by Social Influence \cite{jaques2019social}, AMD \cite{Baumann2020AdaptiveMD}, LIO \cite{LIO} and ID \cite{yang2021adaptive}. Social Influence \cite{jaques2019social} rewards the agent action that has the most impact on others' behavior as an intrinsic reward. In LIO \cite{LIO} an agent learns to use incentive reward that affects the learning update of opponents' policies and changes the objectives of the recipient agents in the direction that improves incentivizer agents' objectives by using meta-gradient learning.
AMD \cite{Baumann2020AdaptiveMD} uses a central planner agent that learns how to set an incentive reward according to agents expected policy update in the next step. \cite{zheng2020ai} presents a two-dimensional grid world dynamic economic environment Gather-Trade-Build game, where agents collect resources, earn coins by building houses with these materials, and trade resources; moreover, there is a central tax-planner agent who learns to improve the trade-off between income equality and productivity by setting taxes that correspond to a payoff from the agent's income.  \cite{yang2021adaptive} use same environment and propose meta-gradient approach to train Incentive Designer (ID), the central planning analogue of LIO, as an incentive mechanism. Mechanism design can also be used to model human incentives and solve human dilemmas such as determining tax rate for a higher social welfare \cite{yang2021adaptive} using the simulation environment AI Economist, proposed in \cite{zheng2020ai}. This is a good illustration of how it can be used as a recommendation system for solving social problems in the future. Because we adopt the approach of directly incentivizing the agents using an extra additive reward and the economic simulation environment from Zheng et al. \cite{zheng2020ai} requires an indirect approach such as determining the tax policy, we do not address the taxation problem in AI Economist in this work and leave it to future work.

Incentivization practices that use meta-gradient learning to shape opponent behavior, such as LIO and ID are the approaches closest to our learning algorithm. However, while in LIO and ID, the meta-gradient based incentive mechanism performs on-policy learning \cite{LIO, yang2021adaptive}, IQ-Flow's incentive mechanism learns in an off-policy manner with a replay buffer. A prior work that uses off-policy learning with a replay buffer for the first time in meta-gradient learning is MetaL \cite{calian2021balancing}. Unlike the opponent shaping based methods, IQ-Flow does not need access or modelling of other agents' parameters. Instead of focusing on how the behavior of agents change, IQ-Flow focuses on rendering cooperative actions in the Nash-Equilibrium for the possible states. Since non-cooperation would incur a loss for all agents, IQ-Flow tasks the agents' to optimize their returns and choose cooperation; IQ-Flow does not keep track of how agents' behavior policies change. This is due to training the incentivizer critic by offline Implicit Q-Learning, which is the key difference from LIO and ID that use online incentivizer training.


\section{Background}\label{background}
In this work, we assume a Partially Observable MDP (POMDP) where $N$ agents learn independently. $\mathcal{S}$ denotes the global state of the environment, $a^{i} \in a$ denote action of i'th agent in joint action $a$, and $i^{-}$ denotes all agent indices except $i$ with index set denoted as $\mathcal{I} = \{0, 1, ..., N-1\}$ . Observation space of agent $i$ is $\mathcal{O}_{i} = \left\{o_{i} | s \in \mathcal{S}, o_{i} = O(s, i)\right\}$ with the observation function $O: \mathcal{S} \times \mathcal{I} \rightarrow \mathbb{R}^{d}$ that maps the observations to the d-dimensional space. State, observation, action and reward at time step $k$ are denoted as $s_{k}$, $o_{k}$, $a_{k}$, $r_{k}$ respectively along with time horizon $T$ and discount factor $\gamma$. We have the transition function of the environment $\mathcal{T}: \mathcal{S} \times \mathcal{A}^{N} \rightarrow \mathcal{P}(S)$ with $\mathcal{P}$ denoting the probability distribution over S and batch length $l_{B}$. Joint reward provided by the environment is $R_{env}: \mathcal{S} \times \mathcal{A}^{N} \rightarrow \mathbb{R}^{N}$ where each agent receives a specific reward $R_{env}^{i}: \mathcal{S} \times \mathcal{A}^{N} \rightarrow \mathbb{R}$. The incentive reward that can be given to an agent is constrained according to the environment as $\mathcal{U} \subset \mathbb{R}$. Thus, the joint incentives provided by the mechanism and parametrized by $\eta$ is $R_{inc, \eta}: \mathcal{S} \times \mathcal{A}^{N} \rightarrow \mathcal{U}^{N} \subset \mathbb{R}^{N}$ where each agent receives a specific incentive $R_{inc, \eta}^{i}: \mathcal{S} \times \mathcal{A}^{N} \rightarrow \mathcal{U} \subset \mathbb{R}$. We define the total reward an agent receives which directs that agent's behavior policy as $R_{ind}^{i} = R_{env}^{i} + R_{inc, \eta}^{i}$. We further define the sum of the rewards that environment provides to all agents as $R_{coop}^{i} = \sum_{id=0}^{N-1}R_{env}^{id}$. It should be noted that $R_{coop}^{i}$ is defined for all agents with the same value. 

We define three different policies that are necessary for our problem case and solution method.
\begin{itemize}
    \item $\boldsymbol\pi_{b}^{i} \in \boldsymbol\pi_{b}$: i'th agent's behavior policy which is optimized to maximise \\ $V_{\boldsymbol\pi_{b}, ind}^{i}(s) := \mathbb{E}_{\boldsymbol\pi_{b}}\left[\sum_{t=k}^{T-1}\gamma^{t-k}R_{ind, t}^{i}|s_k=s\right]$
    \item $\boldsymbol\pi_{coop}^{i} \in \boldsymbol\pi_{coop}$: i'th agent's cooperative policy which is optimized to maximise \\ $V_{\boldsymbol\pi_{coop}}^{i}(s) := \mathbb{E}_{\boldsymbol\pi_{coop}}\left[\sum_{t=k}^{T-1}\gamma^{t-k}R_{coop, t}^{i}|s_k=s\right]$
    \item $\boldsymbol\pi_{env}^{i} \in \boldsymbol\pi_{env}$: i'th agent's environment policy which is optimized to maximise \\ $V_{\boldsymbol\pi_{env}, env}^{i}(s) := \mathbb{E}_{\boldsymbol\pi_{env}}\left[\sum_{t=k}^{T-1}\gamma^{t-k}R_{env, t}^{i}|s_k=s\right]$
\end{itemize}

We further denote the different objectives that are necessary for our problem case and solution method as follows:

\begin{itemize}
    \item Action-values of $i$'th agent under $\boldsymbol\pi_{b}$ accounting for the individual total reward $R_{ind}^{i}$ \\
    $Q_{\boldsymbol\pi_{b}, ind}^{i}(s, a) = \mathbb{E}_{\boldsymbol\pi_{b}}\left[\sum_{t=k}^{T-1}\gamma^{t-k}R_{ind, t}^{i}|s_k=s, a_k=a\right]$
    \item Action-values of $i$'th agent under $\boldsymbol\pi_{coop}$ accounting for the cooperative reward $R_{coop}^{i}$ \\
    $Q_{\boldsymbol\pi_{coop}}^{i}(s, a) = \mathbb{E}_{\boldsymbol\pi_{coop}}\left[\sum_{t=k}^{T-1}\gamma^{t-k}R_{coop, t}^{i}|s_k=s, a_k=a\right]$
    \item Action-values of $i$'th agent under $\boldsymbol\pi_{env}$ accounting for the individual environment reward $R_{env}^{i}$ \\
    $Q_{\boldsymbol\pi_{env}, env}^{i}(s, a) = \mathbb{E}_{\boldsymbol\pi_{env}}\left[\sum_{t=k}^{T-1}\gamma^{t-k}R_{env, t}^{i}|s_k=s, a_k=a\right]$ 
    \item Values of $i$'th agent under $\boldsymbol\pi_{b}$ accounting for the individual environment reward $R_{env}^{i}$ \\
    $V_{\boldsymbol\pi_{b}, env}^{i}(s) = \mathbb{E}_{\boldsymbol\pi_{b}}\left[\sum_{t=k}^{T-1}\gamma^{t-k}R_{env, t}^{i}|s_k=s\right]$
    \item Action-values of $i$'th agent under $\boldsymbol\pi_{b}$ accounting for the individual environment reward $R_{env}^{i}$ \\
    $Q_{\boldsymbol\pi_{b}, env}^{i}(s, a) = \mathbb{E}_{\boldsymbol\pi_{b}}\left[\sum_{t=k}^{T-1}\gamma^{t-k}R_{env, t}^{i}|s_k=s, a_k=a\right]$
    \item Values of $i$'th agent under $\boldsymbol\pi_{b}$ accounting for the individual incentive reward $R_{inc}^{i}$ \\
    $V_{\boldsymbol\pi_{b}, inc}^{i}(s) = \mathbb{E}_{\boldsymbol\pi_{b}}\left[\sum_{t=k}^{T-1}\gamma^{t-k}R_{inc, t}^{i}|s_k=s\right]$ 
    \item Action-values of $i$'th agent under $\boldsymbol\pi_{b}$ accounting for the individual incentive reward $R_{inc}^{i}$ \\
    $Q_{\boldsymbol\pi_{b}, inc}^{i}(s, a) = \mathbb{E}_{\boldsymbol\pi_{b}}\left[\sum_{t=k}^{T-1}\gamma^{t-k}R_{inc, t}^{i}|s_k=s, a_k=a\right]$
\end{itemize}

\paragraph{Social Dilemma conditions}

\begin{table}[h!]
\centering
\caption{Matrix Game payoff table}
\label{tab:mg_payoff}
\begin{tabular}{ c | c c }
  & C & D \\ 
  \hline
 C & R, R & S, T \\  
 \hline
 D & T, S & P, P    
\end{tabular}
\end{table}

According to preliminary work in social dilemmas \cite{learningdynamicsinsd, leibo2017multi}, a Matrix Game such as Table \ref{tab:mg_payoff} is a Social Dilemma if it satisfies the following conditions: 
\begin{enumerate}
    \item $R > P$
    \item $R > S$
    \item $2R > T + S$
    \item $T > R$ or
          $P > S$
\end{enumerate}

In this canonical Matrix Game in Table \ref{tab:mg_payoff} actions C and D represent cooperate and defect actions as convention dictates \cite{learningdynamicsinsd}. We adopt the definitions proposed by \cite{learningdynamicsinsd} and we denote R, P, T and S respectively as reward from mutual cooperation, punishment from mutual defection, temptation reward from defecting while the other player cooperates and sucker reward from cooperating while the other player defects.

\paragraph{Offline Implicit Q-Learning}

Offline Implicit Q-Learning is performed to learn critics as proposed by \cite{implicitqlearning} for dataset $\mathcal{D}$, value parameters $\psi$, critic parameters $\theta$, target critic parameters $\bar{\theta}$, state $s$, action $a$, next state $s^{'}$, discount $\gamma$, expectile $\tau_{exp} \in (0,1)$ with the following loss equations: \\

\begin{equation}
\begin{split}
&L_{2}^{\tau_{exp}}(u)=\left|\tau_{exp}-\mathds{1}(u<0)\right|u^{2} \\
&L_{V}(\psi)=\mathbb{E}_{(s,a)\sim\mathcal{D}}\left[L_2^{\tau_{exp}}(Q_{\bar{\theta}}(s,a)-V_{\psi}(s))\right] \\
&L_{Q}(\theta)=\mathbb{E}_{(s,a,s^{'})\sim\mathcal{D}}\left[(r(s,a)+\gamma V_{\psi}(s^{'})-Q_{\theta}(s,a))^{2}\right] \text{\cite{implicitqlearning}}
\end{split}
\end{equation}

We extend offline Implicit Q-Learning to our multi-agent case in order to approximate $Q_{\boldsymbol\pi_{b}, ind}^{i}$, $Q_{\boldsymbol\pi_{coop}}^{i}$, and $Q_{\boldsymbol\pi_{env}, env}^{i}$. We give the corresponding losses in Appendix \ref{marliql}. We denote the training batch with $\mathcal{B}_{T}$ and validation batch $\mathcal{B}_{V}$.


\section{Incentive Q-Flow}

IQ-Flow bases itself on reversing the fourth social dilemma condition and make $T < R$ and $P < S$ in Table \ref{tab:mg_payoff}. When $R > T$ and $S > P$, choosing $C$ over $D$ becomes the greedy policy automatically without regard to the opponents' policy. Thus, IQ-Flow aims to make the action of the cooperative policy the greedy choice for the incentivized behavior policy using meta-gradients as we defined in background in section \ref{background}.

The necessity of using meta-gradients for estimating how Q-Values change according to $\eta$ comes from the fact that it is not possible to directly estimate the long term value change as a result of a change of incentives. Let the optimal actions of the cooperative policy and incentivized behavior policy be defined respectively as:

\begin{equation}
\begin{split}
&a^{i}_{coop} = \argmax_{a^{i}} Q_{\boldsymbol\pi_{coop}}^{i}(s, a^{i^{-}}, .) \\
&a^{i}_{b} = \argmax_{a^{i}} Q_{\boldsymbol\pi_{b}, ind}^{i}(s, a^{i^{-}}, .) \\
\end{split}
\end{equation}

Let the optimal actions for the self-interested policy of agents under standard environment conditions with no extra incentives be defined as:

\begin{equation}
a^{i}_{env} = \argmax_{a^{i}} Q_{\boldsymbol\pi_{env}, env}^{i}(s, a^{i^{-}}, .)
\end{equation}

In order to determine $a^{i}_{coop}$, $a^{i}_{b}$, and $a^{i}_{env}$, IQ-Flow needs to estimate $Q_{\boldsymbol\pi_{coop}}^{i}$, $Q_{\boldsymbol\pi_{b}, ind}^{i}$, and $Q_{\boldsymbol\pi_{env}, env}^{i}$. IQ-Flow approximates $Q_{\boldsymbol\pi_{coop}}^{i}$, $Q_{\boldsymbol\pi_{b}, ind}^{i}$, and $Q_{\boldsymbol\pi_{env}, env}^{i}$ respectively by $Q_{\boldsymbol\pi_{coop}}^{i}\left(\theta_{coop}\right)$, \\ $Q_{\boldsymbol\pi_{b}, ind}^{i}\left(\theta_{ind}\right)$, and $Q_{\boldsymbol\pi_{env}, env}^{i}(\theta_{env})$. An important point is that since incentive function is dynamic, $Q_{\boldsymbol\pi_{b}, ind}^{i}(\theta_{ind})$ and \\ $V_{\boldsymbol\pi_{b}, ind}^{i}(\psi_{ind})$ need to be updated with the $r_{inc}^{i}$ inferred from the last $\eta$. IQ-Flow updates the critic parameters $\psi_{ind}$ and $\theta_{ind}$, respectively for $V_{\boldsymbol\pi_{b}, ind}^{i}(s, \psi_{ind})$ and $Q_{\boldsymbol\pi_{b}, ind}^{i}(s, a, \theta_{ind})$, with Implicit Q-Learning extended to MARL with the equations in Appendix \ref{marliql}.

In order to update $\eta$, we first update our predefined critics with learning rate $\beta_{ind}$ for $K$ steps. This update can be given as following for the Stochastic Gradient Descent (SGD) optimizer:

\begin{equation}
\begin{split}
&\hat{\theta}_{ind} \leftarrow \theta_{ind} + \beta_{ind} \nabla_{\theta_{ind}}\cfrac{1}{l_{B} N}\mathlarger{\sum_{k=0}^{l_{B}-1}}\mathlarger{\sum_{i=0}^{N-1}}\\
&\left(r_{env}^{i}\left(s_{k},a_{k}\right)+r_{inc}^{i}\left(s_{k},a_{k},\eta\right)+\gamma V^{i}_{\psi_{ind}}(s^{'}_{k})-Q^{i}_{\theta_{ind}}\left(s_{k},a_{k}\right)\right)^{2}\\
\end{split}
\end{equation}

Since we want to update $\eta$ in the direction that flows Q-Values from actions of defective policies to actions of cooperative policies, we regard the $a_{coop}$ as target labels in a classification problem and use a modified version of cross-entropy loss. The necessity of the modification in the cross-entropy loss is because we only want the gradient flow as long as there is a dilemma in the system so that there is no unnecessary and excessive incentivization. We identify an action that causes a dilemma as $a^{i}_{b} \neq a^{i}_{coop}$. Therefore we further mask our meta-loss for the case when there is no estimated dilemma. In order to get a probabilistic view of Q-Values and use them in the cross-entropy loss, we pass them through a softmax layer.

Finally our meta-loss can be defined as follows:

\begin{equation}\label{metaloss}
\begin{split}
&L^{m}_{\eta}(\hat{\theta}_{ind}):=-\cfrac{1}{l_{B} N}\mathlarger{\sum_{k=0}^{l_{B}-1}} \mathlarger{\sum_{i=0}^{N-1}}\mathlarger{\sum_{\tilde{a}=0}^{|A|-1}}\mathds{1}\left(\tilde{a}=a^{i}_{coop, k}\right)\\
& \times \left(1 - \mathds{1}(a^{i}_{b, k}=a^{i}_{coop, k})\right)\log\left(\sigma\left(Q_{\boldsymbol\pi_{b}, ind}^{i}\left(s_{k}, a^{i}, a_{k}^{i^{-}},\hat{\theta}_{ind}\right)\right)\right)\Big|_{a^{i}=\tilde{a}}\\
&\sigma(z_{i}) = \mathlarger{\frac{e^{z_{i}}}{\sum_{j}e^{z_{j}}}}\\
\end{split}
\end{equation}

Since we do not want to give an unnecessary incentive if there is no dilemma in the original case without extra incentives, we use another mask which determines if $a^{i}_{env} = a^{i}_{coop}$. Therefore we add a cost regularization term to the meta loss with cost coefficient $c_1$.

\begin{equation}
\begin{split}
&L^{cost_{1}}_{\eta}(\hat{\theta}_{ind}) := \cfrac{1}{l_{B} N} \mathlarger{\sum_{k=0}^{l_{B}-1}} \mathlarger{\sum_{i=0}^{N-1}} \mathlarger{\sum_{act=0}^{|A|-1}}\mathds{1}(a^{i}_{coop, k}=a^{i}_{env, k})\\
&\times \left|Q_{\boldsymbol\pi_{b}, ind}^{i}(s_{k}, a^{i}, a_{k}^{i^{-}} \hat{\theta}_{ind})))\right|\Big|_{a^{i}={act}}\\
\end{split}
\end{equation}

If the incentives become too high prematurely, they can have a destructive effect, especially if they are the wrong incentives. Therefore we add another cost regularization term to the meta loss with cost coefficient $c_2$. Although our experiments show that these cost regularization terms are not required to get a successful performance, especially in simple problems, we find that including them leads to higher performance.

\begin{equation}
\begin{split}
&L^{cost_{2}}_{\eta}(\hat{\theta}_{ind}) := \cfrac{1}{l_{B} N} \mathlarger{\sum_{k=0}^{l_{B}-1}} \mathlarger{\sum_{i=0}^{N-1}} \mathlarger{\sum_{act=0}^{|A|-1}}\left(1 - \mathds{1}(a^{i}_{coop, k}=a^{i}_{env, k})\right)\\
&\times \left|Q_{\boldsymbol\pi_{b}, ind}^{i}(s_{k}, a^{i}, a_{k}^{i^{-}} \hat{\theta}_{ind})))\right|\Big|_{a^{i}={act}}\\
\end{split}
\end{equation}

Our final incentive loss for $\eta$ is given below as $L^{R_{inc}}_{\eta}(\hat{\theta}_{ind})$:

\begin{equation}
L^{R_{inc}}_{\eta}(\hat{\theta}_{ind}) = L^{m}_{\eta}(\hat{\theta}_{ind}) + c_{1} L^{cost_{1}}_{\eta}(\hat{\theta}_{ind}) + c_{2} L^{cost_{2}}_{\eta}(\hat{\theta}_{ind})
\end{equation}

If we use $\alpha$ as learning rate for $\eta$, set number of critic update steps $K$ as 1, and assume SGD for optimizer, the update becomes:

\begin{equation}\label{fullmetaloss}
\begin{split}
&\hat{\eta} \leftarrow \eta + \alpha \nabla_{\eta}L^{R_{inc}}_{\eta}(\hat{\theta}_{ind})\\
&\nabla_{\eta}L^{R_{inc}}_{\eta}(\hat{\theta}_{ind}) = \dfrac{\partial L^{m}_{\eta}(\hat{\theta}_{ind}) + c_{1} L^{cost_{1}}_{\eta}(\hat{\theta}_{ind}) + c_{2} L^{cost_{2}}_{\eta}(\hat{\theta}_{ind})}{\partial \hat{\theta}_{ind}} \dfrac{\partial \hat{\theta}_{ind}}{\partial \eta}\\
\end{split}
\end{equation}

The diagram for how $\eta$ meta-parameter is updated is given below in Figure \ref{fig:diag}:

\begin{figure}[H]
\centering
    \includegraphics[scale=0.3]{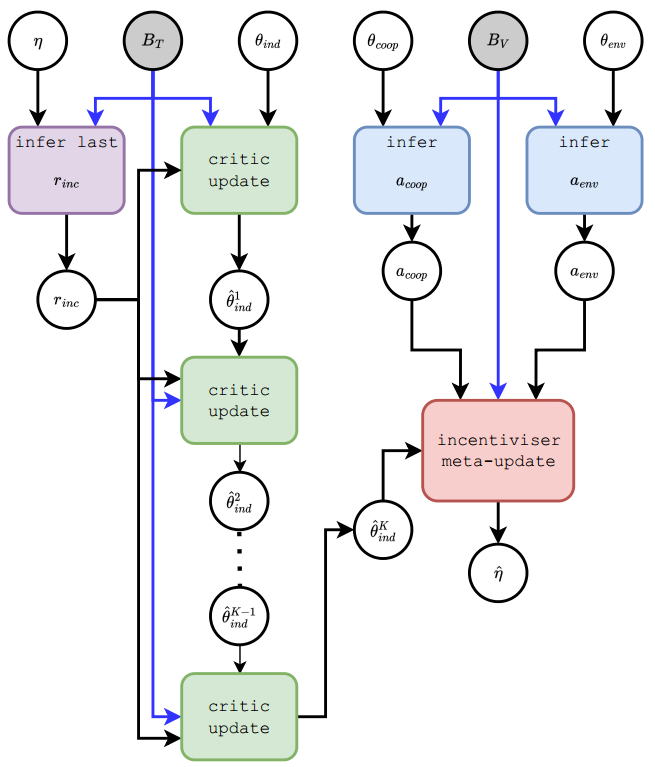}
\caption{Meta-update diagram for incentive parameter $\eta$}
\label{fig:diag}
\end{figure}

The pseudocode of the algorithm is given below in Algorithm \ref{alg:iqflowalgo}.
\begin{algorithm}[H]
\caption{Incentive Q-Flow}\label{alg:iqflowalgo}
\begin{algorithmic}
\Procedure{Train IQ-Flow Mechanism}{$\phi^{0}$, $\phi^{1}$,...,$\phi^{N-1}$, args} \Comment{Input: policy of all agents, hyperparameters}
\State Initialize $\eta$, $\theta_{coop}$, $\theta_{env}$, $\theta_{ind}$, $\psi_{coop}$, $\psi_{env}$, $\psi_{ind}$
\State $num\_episode \leftarrow 0$ 
\For{number of episodes to train}
\State Run agents with policies $\phi^{0}$, $\phi^{1}$,...,$\phi^{N-1}$ for an episode with incentives given by $\eta$
\State $num\_episode \leftarrow num\_episode + 1$ 
\State Add the transitions from episode to replay buffer of IQ-Flow
\State Update agent policies $\phi^{0}$, $\phi^{1}$,...,$\phi^{N-1}$ according to their private learning rules
\State Update $\theta_{coop}$, $\theta_{env}$, $\theta_{ind}$, $\psi_{coop}$, $\psi_{env}$, $\psi_{inc}$ using equations in \ref{criticupdate}
\State sample $\mathcal{B}_{T}$ and $\mathcal{B}_{V}$ for meta-update
\State simulate mechanism critic update for $K$ times using $\mathcal{B}_{T}$, $\theta_{ind}$
\State Update $\eta$ using $\mathcal{B}_{V}$ (with equations \ref{metaloss} or \ref{fullmetaloss})
\EndFor
\EndProcedure
\end{algorithmic}
\end{algorithm}


\section{Experiments}

\subsection{Iterated Matrix Games}

\begin{figure}[H]
\centering
    \includegraphics[scale=0.19]{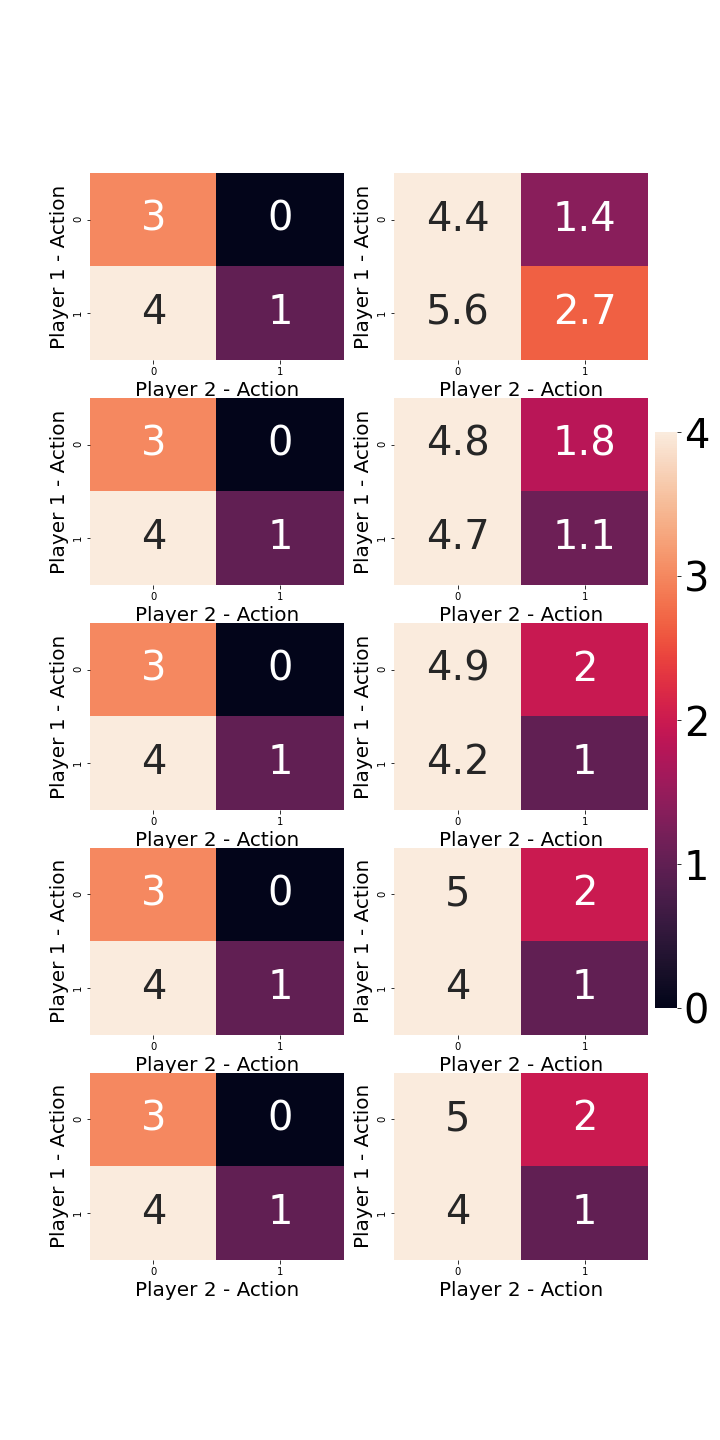}
\caption{IPD player 1 payoff matrices}
\label{p1ipdpayoff}
\end{figure}

We demonstrate how IQ-Flow operates on the iterated extension of the three canonical Matrix Games, which are Prisoner's Dilemma, Chicken Game, and Stag Hunt. The payoff matrices for these games are given in Table \ref{tab:pd}, Table \ref{tab:chick}, and Table \ref{tab:stag}. We extend the implementation used by LOLA \cite{foerster2017learning} and use the policy gradient agents for the independent learners as used by LIO \cite{LIO} and ID \cite{yang2021adaptive}. The incentive reward is set as $R^{i}_{inc} \in (0, 2)$ to provide only sufficient incentivization and number of iterations is set as 20 for all experiments. Since the experimentation purpose here is for illustration rather than comparison, hyperparameter tuning was not performed to optimize learning performance and cost regularization was not added to the meta-objective. We demonstrate how IQ-Flow changes the payoff matrix of the games in Figure \ref{p1ipdpayoff} and Appendix \ref{ipdfull}. The first column in Figure \ref{p1ipdpayoff} represents the original payoffs. The other column represents the modified total payoffs by IQ-Flow where each row represents the mechanism state trained for 30, 210, 390, 570, 750 episodes respectively. The first rows in the figures in Appendix \ref{ipdfull} represent the original payoffs, while the other rows represent the state (initial state, previous action taken CC, previous action taken CD, previous action taken DC, and previous action taken DD). The columns represent the total payoff output of the  mechanism state trained for 30, 210, 390, 570, 750 episodes respectively.

\begin{figure}[h]
\centering
    \includegraphics[scale=0.19]{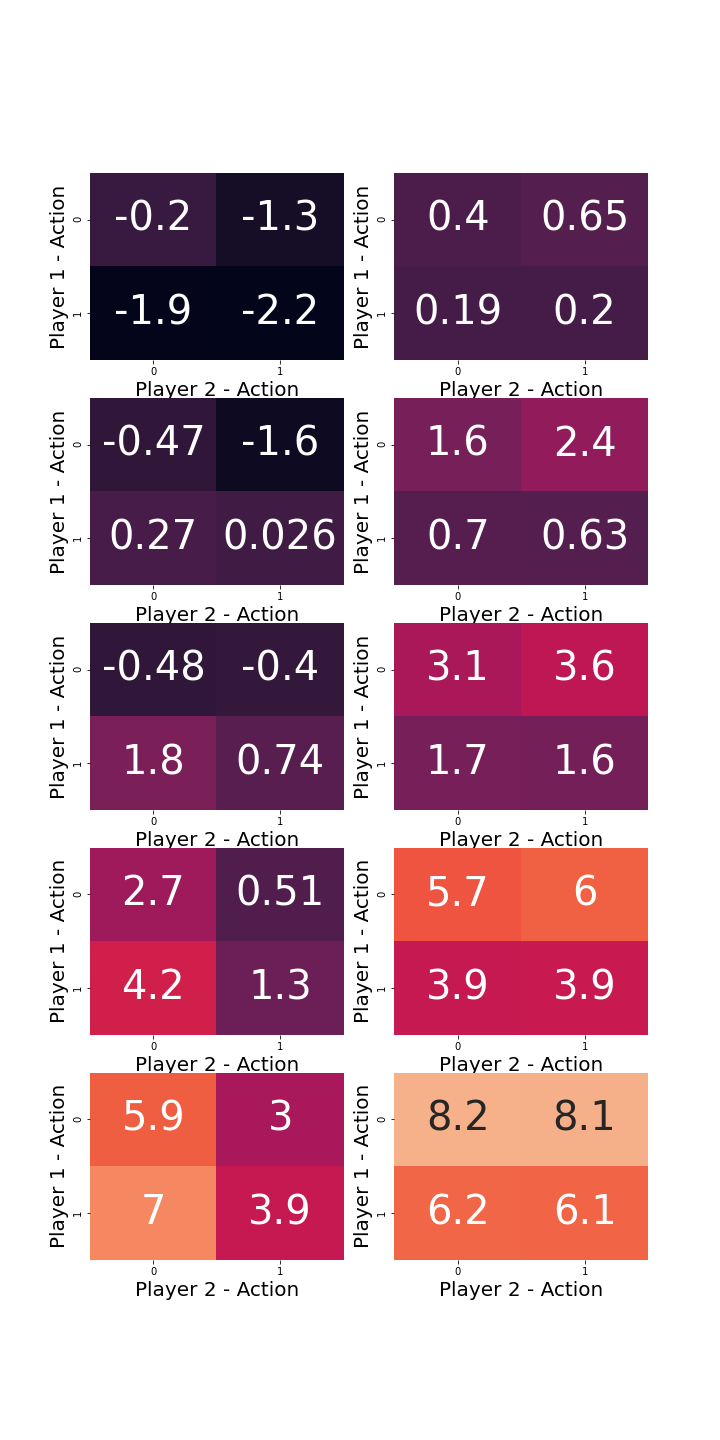}
\caption{IPD player 1 estimated Q-Value matrices (left: without incentives, right: with incentives)}
\label{p1ipdpayoffqtable}
\end{figure}

We depict how IQ-Flow changes the estimated Q-Values of the games in Figure \ref{p1ipdpayoffqtable} and Appendix \ref{ipdfullqtable}. The first columns in Figure \ref{p1ipdpayoffqtable} and figures in Appendix \ref{ipdfullqtable} represent the Q-Values without the mechanism incentives. The other columns represent the Q-Values with the mechanism incentives where each row represents the mechanism state trained for 30, 210, 390, 570, 750 episodes respectively. These outputs are given for the initial state.

\begin{table}[ht]
    \begin{minipage}{.5\linewidth}
    \centering
    \caption{Prisoner's Dilemma}
    \label{tab:pd}
    \begin{tabular}{ c | c c }
    PD & $C_{2}$ & $D_{2}$ \\ 
      \hline
     $C_{1}$ & (3, 3) & (0, 4) \\  
     \hline
     $D_{1}$ & (4, 0) & (1, 1)    
    \end{tabular}
    \end{minipage}
    \begin{minipage}{.5\linewidth}
    \centering
    \caption{Chicken Game}
    \label{tab:chick}
    \begin{tabular}{ c | c c }
    Chicken & $C_{2}$ & $D_{2}$ \\ 
      \hline
     $C_{1}$ & (3, 3) & (1, 4) \\  
     \hline
     $D_{1}$ & (4, 1) & (0, 0)    
    \end{tabular}
    \end{minipage}
    \begin{minipage}{.5\linewidth}
    \centering
    \caption{Stag Hunt}
    \label{tab:stag}
    \begin{tabular}{ c | c c }
    Stag Hunt & $C_{2}$ & $D_{2}$ \\ 
      \hline
     $C_{1}$ & (4, 4) & (0, 3) \\  
     \hline
     $D_{1}$ & (3, 0) & (1, 1)    
    \end{tabular}
    \end{minipage}
\end{table}

\begin{figure}[H]
\centering
    \includegraphics[scale=0.22]{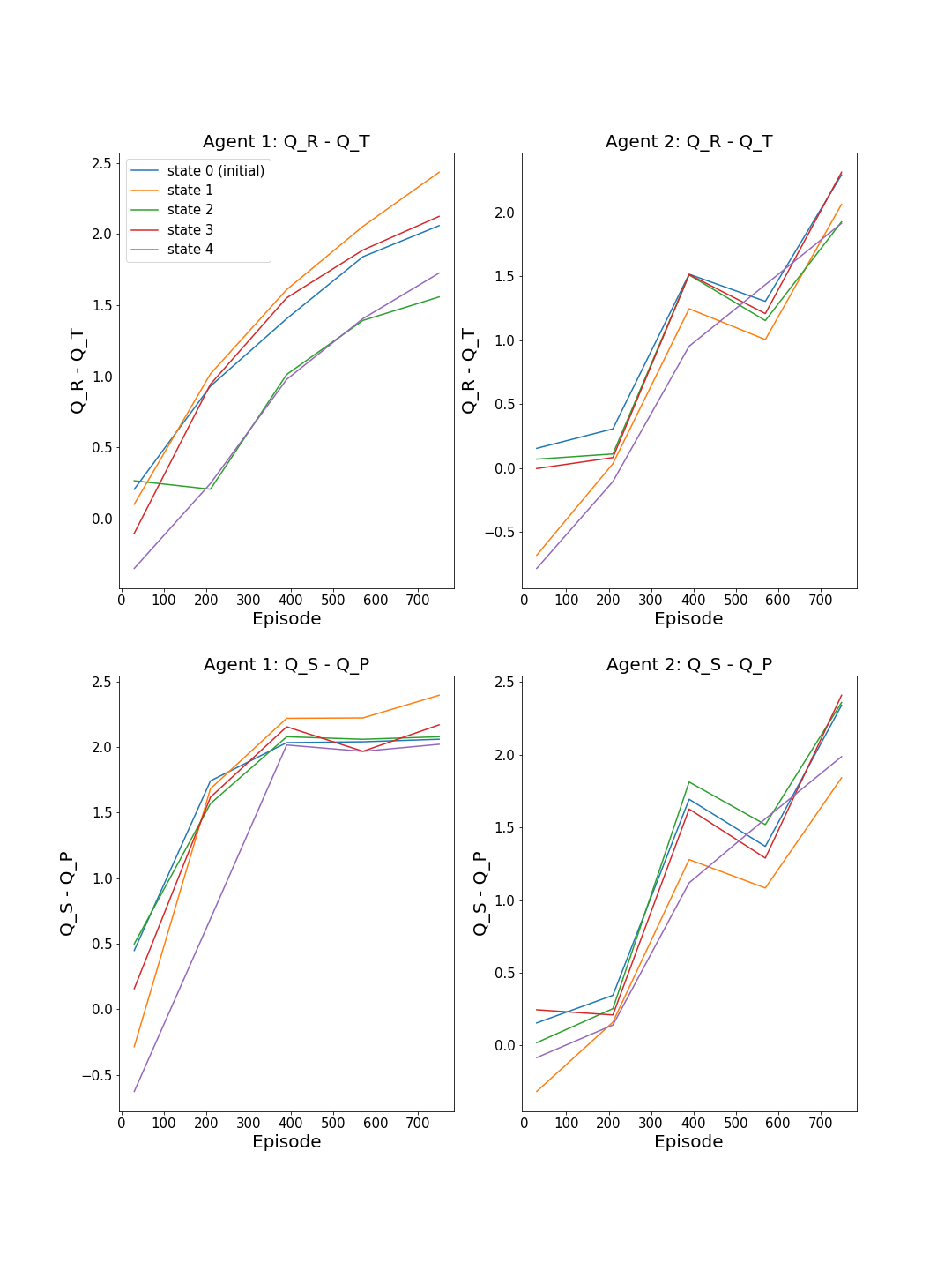}
\caption{IPD $R - T$ and $S - P$ plot for Q-Values}
\label{p1ipdineq}
\end{figure}

In addition to the detailed payoff and Q-Value charts, we provide a plot for Iterated Matrix Games to show how the inequalities turn from $T > R$ and/or $P > S$ to $T < R$ and $P < S$ in Figure \ref{p1ipdineq} and Appendix \ref{ipdineq} as training progresses. We highlight that the $R$, $T$, $P$, and $S$ denotes the corresponding estimated Q-Values for all states and not the single step payoffs.

Consequently, our results demonstrate clearly that IQ-Flow is capable of removing the social dilemma for Iterated Prisoner's Dilemma, Chicken Game and Stag Hunt; since we obtain $T < R$ and $P < S$ in the end for all of the cases in both single step payoffs and estimated future returns.

\subsection{Escape Room}

Escape room is a small, N-player Markov game proposed by \cite{LIO}. The game contains 3 different state: initial, lever and door state where agents spawn in the initial state and aim to reach the door which is the terminal state \cite{LIO}. But M number of agents must cooperate by pulling the lever at the same time to get others out of the door so that the agent who goes out of the door gets +10 reward individually while the cost of the pulling lever is -1 \cite{LIO}. Therefore, in order to increase total return, some of the agents should give up their own interest and act cooperatively. We extend the implementation used by LIO \cite{LIO}, benefit from \cite{kostrikovjaxrl}, and use the policy gradient agents for the independent learners as used by LIO \cite{LIO} and ID \cite{yang2021adaptive}. We use the same experiment setup used by ID \cite{yang2021adaptive} and evaluate IQ-Flow's performance along with an ablation study given in Appendix \ref{ablationextra}.

\begin{figure}[H]
  \centering
  \includegraphics[width=\linewidth]{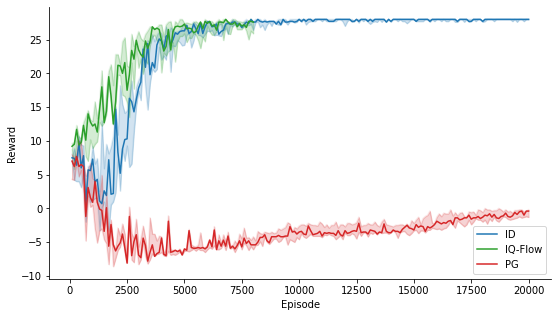}
  \caption{ER(5,2)}
  \label{fig:erres1}
\end{figure}

\begin{figure}[H]
  \centering
  \includegraphics[width=\linewidth]{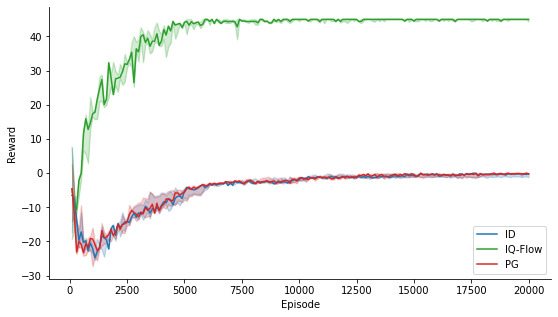}
  \caption{ER(10,5)}
  \label{fig:erres2}
\end{figure}

We give the results of Escape Room (5, 2) experiment, as in \cite{yang2021adaptive}, in Figure \ref{fig:erres1}. The basic case of no incentivization, denoted as PG, performs poorly as expected. ID reaches the optimal total return of the environment, which is 28. IQ-Flow performs the best by reaching 28 faster and with better initial training performance. The results of the experiment Escape Room (10, 5) is given in Figure \ref{fig:erres2}.  The basic case of no incentivization, denoted as PG, performs poorly again as expected. Although Yang et al. show \cite{yang2021adaptive} that ID reaches the optimal return of 45, we could not replicate those results with our implementation and obtained the performance of ID similar to PG. IQ-Flow reaches the optimal return of 45 faster than ID in both our implementation and the resuls given in \cite{yang2021adaptive}. Since the results of ablation study, given in Appendix \ref{ablationextra}, does not provide distinctive results, we focus on the ablation of experiments in the 2-Player Cleanup environment.

\subsection{SSD Environment - 2-Player Cleanup}

Cleanup \cite{hughes2018inequity} is a grid-world social dilemma environment where the objective is to collect apples from field that give +1 reward. Since the respawn time of the apples depends on the amount of waste, which increases over time, if the amount of waste exceeds a threshold no apples can spawn \cite{hughes2018inequity}; therefore, agents need to clean the waste by using clean beam skills for apples to continue to spawn even though staying in the apple field returns more individual rewards. We use decentralized independent actor critic learners and the same environment setup with 2 agents, which we call the 2-Player Cleanup environment, as used by LIO \cite{LIO} and ID \cite{yang2021adaptive} for the $7\times7$ map.

\begin{figure}[H]
  \centering
  \includegraphics[width=\linewidth]{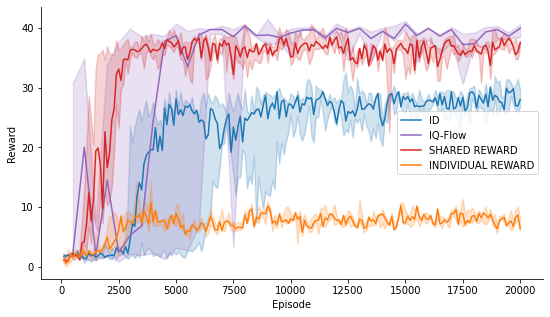}
  \caption{$7\times7$ experiment result}
  \label{fig:clres1}
\end{figure}

It can be seen from Figure \ref{fig:clres1} that IQ-Flow performs the best while reaching the return upper bound, identified by shared reward agent's performance as in LIO \cite{LIO}. Decentralized actor critic agents perform poor as expected while the decentralized actor critic agents with the shared centralized reward set the return upper bound. Although ID performs close to the return upper bound in both our implementation and the results provided by Yang et al. in \cite{yang2021adaptive}, it fails to reach it. It should also be noted that while IQ-Flow performs best and reaches the upper bound for good runs, it has high variance close to the end of training for naive training. This variance occurs due to some loss in performance when the actor critic agents' policies get too disconnected from the mechanism. Therefore, in order to obtain a stable training, we reset the actor-critic agents in the environment each 1000 episodes. Since after each reset operation the actor-critic agents start learning from scratch, we sample evaluation results each 500 episodes in order to filter the pseudo-loss in performance caused by learning from scratch and have a fair comparison with other algorithms.

\begin{figure}[H]
  \centering
  \includegraphics[width=\linewidth]{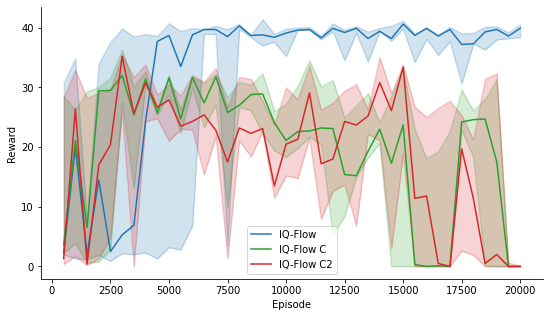}
  \caption{Ablation results}
  \label{fig:clres2}
\end{figure}

The ablation results for 2-Player Cleanup is given in Figure \ref{fig:clres2}. IQ-Flow denotes the standard algorithm with cost regularization cost 1 and cost 2. IQ-Flow C denotes the case when cost coefficient 1 is 0 and IQ-Flow C2 denotes the case where there is no cost regularization. It is demonstrated that having cost regularization with both coefficients greater than 0 indeed increases learning performance.

\begin{figure}[H]
\centering
    \includegraphics[width=\linewidth]{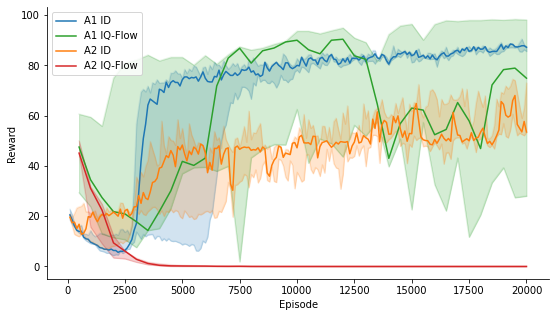}
\caption{Incentive Rewards given by IQ-Flow and ID}
\label{ssdincentive}
\end{figure}

The incentive rewards provided by ID and IQ-Flow in 2-Player Cleanup Environment is given in Figure \ref{ssdincentive}.

The incentive rewards given by IQ-Flow and ID to agent 1 and agent 2 are presented in Figure \ref{ssdincentive}. Incentive rewards given to agent 1 (A1, cleaner) are in close range with each other for IQ-Flow and ID, but incentive rewards given to agent 2 (A2, harvester) are dissimilar. While ID learns to give an unnecessary incentive to the harvester agent, IQ-Flow learns not to give any unnecessary incentive to this harvester agent. This is attributed to IQ-Flow's capacity to infer when there is a dilemma and when there is no dilemma.

\begin{figure}[H]
\centering
    \includegraphics[width=\linewidth]{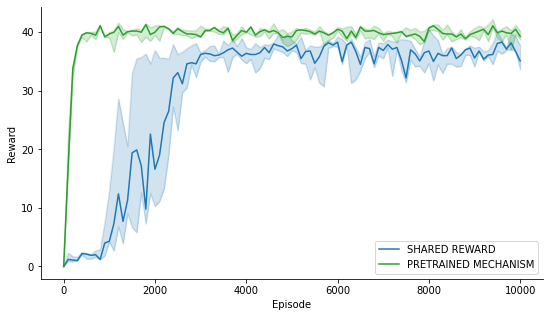}
\caption{Comparison between pretrained IQ-Flow mechanism and shared reward setup}
\label{pretrain}
\end{figure}

Finally, we demonstrate in Figure \ref{pretrain} how a reward system supported by a pretrained incentive mechanism by IQ-Flow performs in comparison to a shared reward system. Although the shared reward case with actor-critic agents gives the return upper-bound for 2-Player Cleanup environment, incentivized case with pretrained and frozen IQ-Flow mechanism and actor-critic agents yields much faster learning with higher performance.


\section{Conclusion} \label{conc}
In conclusion, we presented a new algorithm named IQ-Flow to design incentivizers to remove a social dilemma from an environment without any need to perform opponent modelling or access to internal agent parameters. IQ-Flow is fully decentralized and uses the offline RL method Implicit Q-Learning to evaluate policies which are not available in the experienced data. We demonstrated how IQ-Flow modifies the payoff matrix and estimated Q-Values of Iterated Matrix Games for both players, and that it outperforms ID in the existing sequential social dilemma benchmarks. We also demonstrated how much more efficient the reward setup that IQ-Flow produces is than the shared reward case.  We consider a promising direction for future work in this area to learn incentive designers with IQ-Flow from offline data with fully offline training so that we can have a method to remove dilemmas from real world that we can not simulate.



\begin{acks}
We thank Tolga Ok for the valuable discussions throughout this research. Bengisu Guresti thanks the DeepMind scholarship program for the support during her studies. This work is supported by the Scientific Research Project Unit (BAP) of Istanbul Technical University, Project Number: MOA-2019-42321.

\end{acks}



\bibliographystyle{ACM-Reference-Format} 
\bibliography{ref}


\clearpage
\appendix

\section{Appendix} \label{appx}

\subsection{MARL Critic Losses by Extending Implicit Q-Learning}\label{marliql}

\begin{equation}\label{criticupdate}
\begin{split}
& Loss_{V_{\boldsymbol\pi_{coop}}^{i}}(\psi_{coop})=\cfrac{1}{l_{B} N} \mathlarger{\sum_{k=0}^{l_{B}-1}} \mathlarger{\sum_{i=0}^{N-1}} L_2^{\tau_{exp}} \Big( Q_{\bar{\theta}_{coop}}^{i}(s_{k},a_{k}) \\
& - V_{\psi_{coop}}^{i}(s_{k}) \Big) \\
& Loss_{Q_{\boldsymbol\pi_{coop}}^{i}}(\theta_{coop})=\cfrac{1}{l_{B} N}\mathlarger{\sum_{k=0}^{l_{B}-1}} \mathlarger{\sum_{i=0}^{N-1}}\Big(\sum_{j=0}^{N-1}r_{env}^{j}(s_{k},a_{k}) \\
& + \gamma V^{i}_{\psi_{coop}}(s^{'}_{k})-Q^{i}_{\theta_{coop}}(s_{k},a_{k})\Big)^{2} \\
& Loss_{V_{\boldsymbol\pi_{b}, ind}^{i}}(\psi_{ind})=\cfrac{1}{l_{B} N}\mathlarger{\sum_{k=0}^{l_{B}-1}} \mathlarger{\sum_{i=0}^{N-1}}L_2^{\tau_{exp}}\Big(Q_{\bar{\theta}_{ind}}^{i}(s_{k},a_{k}) \\
& - V_{\psi_{ind}}^{i}(s_{k})\Big) \\
& Loss_{Q_{\boldsymbol\pi_{b}, ind}^{i}}(\theta_{ind})=\cfrac{1}{l_{B} N}\mathlarger{\sum_{k=0}^{l_{B}-1}} \mathlarger{\sum_{i=0}^{N-1}}\Big(r_{ind}^{i}(s_{k},a_{k}) \\
& + \gamma V^{i}_{\psi_{ind}}(s^{'}_{k})-Q^{i}_{\theta_{ind}}(s_{k},a_{k})\Big)^{2} \\
& Loss_{V_{\boldsymbol\pi_{env}, env}^{i}}(\psi_{env})=\cfrac{1}{l_{B} N}\mathlarger{\sum_{k=0}^{l_{B}-1}} \mathlarger{\sum_{i=0}^{N-1}}L_2^{\tau_{exp}}\Big(Q_{\bar{\theta}_{env}}^{i}(s_{k},a_{k}) \\
& - V_{\psi_{env}}^{i}(s_{k})\Big)\\
& Loss_{Q_{\boldsymbol\pi_{env}, env}^{i}}(\theta_{env})=\cfrac{1}{l_{B} N}\mathlarger{\sum_{k=0}^{l_{B}-1}} \mathlarger{\sum_{i=0}^{N-1}}\Big(r_{env}^{i}(s_{k},a_{k}) \\
& + \gamma V^{i}_{\psi_{env}}(s^{'}_{k})-Q^{i}_{\theta_{env}}(s_{k},a_{k})\Big)^{2}\\
\end{split}
\end{equation}

\subsection{Implementation Details}

In IQ-Flow experiments, the agents are equipped with Actor-Critic networks that have a similar structure with LIO \cite{LIO} and ID \cite{yang2021adaptive}. The input of the agents' actor network is an image, which passes through a single convolution layer with six filters of size [3,3]. The output of the convolution operation is flattened and fed to three consecutive fully connected layers with sizes 64,64,6 respectively. Rectified Linear Unit (ReLU) activation function is applied in all layers except the last layer, where the softmax activation function is used to produce the activation probabilities.
 
The critic network shares the same structure as the actor network up until the last layer. However, in the critic network, the last layer has a size of 1 and does not apply any activation function. 

The IQ-Flow algorithm employs seven neural networks, excluding target critics, each designed with a specific objective. These networks include Q-value (state-action value), V-value (state value), and Incentive (Rewarder) networks of different types. The objectives of these networks are cooperative reward Q-value, cooperative reward V-value, environment reward Q-value, environment reward V-value, incentive reward Q-value, incentive reward V-value, and incentive reward, respectively. To ensure an efficient training, target critic networks are employed for each Q-value network in accordance with Implicit Q-Learning \cite{implicitqlearning}. During training, after updating the agent policies, we perform $K=20$ critic updates for the meta-update process.

To ensure consistency in state input across environments that do not provide a global state, we implemented a concatenation operation to assume the state as the union of agent observations. When the observations are in the form of 1-D vectors, the concatenation operation is performed first and the resulting concatenated vector is then fed to the networks. For 2-D image observations, all image inputs are first passed through a shared convolution layer, and the resulting embedding vectors are then concatenated to obtain the state more efficiently.

In the case of Q-Value networks and Incentive networks, the assumed state embeddings are concatenated with the union of the agents' actions in 1-hot form for the full action space, where each agent's own actions are masked. We use the full action space for this operation in all experiments of ID and IQ-Flow, as opposed to Yang et al. \cite{yang2021adaptive}, where the ID algorithm uses binary input indicating whether a cleaning beam was used or not in the Cleanup environment for their Incentive network.

The hyperparameters used in both the implementation and experiments are listed in \ref{hypers}.

\subsection{Experimental Setup} \label{expset}

We implement IQ-Flow and ID \cite{LIO} algorithm with JAX \cite{jax2018github, flax2020github, kostrikovjaxrl} which automatically differentiate through functions, loops, and other operations. 

All experiments were run on the following hardware: 2x28 cores Intel(R) Xeon(R) Gold 6258R CPU @ 2.70GHz, with 4 Nvidia Geforce 2080 Ti graphic cards. 

\subsection{Hyperparameters} \label{hypers}

Random exploration rate epsilon is denoted by $\epsilon$, reward discount factor is denoted by $\gamma$, policy entropy coefficient is denoted by $c_{entropy}$, learning rate of the actor network is denoted by $lr_{actor}$, incentive network learning rate is denoted by $lr_{reward}$, cost regularization coefficients are denoted by $c_{costreg}$ and $c_{costreg2}$, fully connected layer size is denoted by $h$, and rewarder network fully connected layer size is denoted by $hr$.
 
 \begin{table}[h!]
\centering
\caption{PG hyperparameters in Escape Room}
\label{tab:PGER}
\begin{tabular}{ c | c c }
  Parameter & \textbf{ER(5,2)} & \textbf{ER(10,5)} \\ 
  \hline
 $\gamma$ & 0.99 & 0.99\\
 $lr_{actor}$ & $9.56e^{-5}$ & $9.56e^{-5}$\\
 $c_{entropy}$ & $1.66e^{-2}$ & $1.66e^{-2}$\\  
 $h_{1}$ & $64$ & $64$\\  
 $h_{2}$ & $64$ & $64$\\
 \hline
\end{tabular}
\end{table}

\begin{table}[h!]
\centering
\caption{ID hyperparameters in Escape Room}
\label{tab:IDER}
\begin{tabular}{ c | c c }
  Parameter & \textbf{ER(5,2)} & \textbf{ER(10,5)} \\ 
  \hline
 $\epsilon_{end}$ & 0.05 & 0.5\\
 $\epsilon_{start}$ & 1.0 & 1.0\\
 $\gamma$ & 0.99 & 0.99\\
 $lr_{actor}$ & $9.56e^{-5}$ & $9.56e^{-5}$\\
 $lr_{cost}$ & $6.03e^{-5}$ & $6.03e^{-5}$\\
 $c_{entropy}$ & $1.66e^{-2}$ & $1.66e^{-2}$\\  
 $lr_{reward}$ & $7.93e^{-4}$ & $7.93e^{-4}$\\
 $c_{reg}$ & $1.0$ &  1.0 \\
 $h_{1}$ & $64$ & $64$\\  
 $h_{2}$ & $64$ & $64$\\
 $hr_{1}$ & $64$ & $64$\\  
 $hr_{2}$ & $32$ & $32$\\
 \hline
\end{tabular}
\end{table}

\begin{table}[h!]
\centering
\caption{ID hyperparameters in 2-Player Cleanup}
\label{tab:IDSSD}
\begin{tabular}{ c | c}
  Parameter & \textbf{}\\ 
  \hline
 $\epsilon_{end}$ & 0.05 \\
 $\epsilon_{start}$ & 1.0 \\
 $\gamma$ & 0.99 \\
 $c_{entropy}$ & 0.43554\\
 $lr_{actor}$ & $1.7841e^{-5}$\\
 $lr_{reward}$ & $4.1990e^{-5}$\\
 $lr_{v}$ & $3.05892e^{-5}$\\
 $lr_{vmodel}$ & $2.1070e^{-5}$\\
 $c_{reg}$ & 1e-4\\
 $\tau$ & 0.1 \\
 $h_{1}$ & $64$\\  
 $h_{2}$ & $64$\\
 $filters$ & $6$\\
 $kernel$ & $[3, 3]$\\
 \hline
\end{tabular}
\end{table}

\begin{table}[h!]
\centering
\caption{IQ-Flow hyperparameters in Prisoner’s Dilemma}
\label{tab:IQIPD}
\begin{tabular}{ c | c}
  Parameter & \textbf{} \\ 
  \hline
 $\epsilon_{end}$ & 0.05 \\
 $\epsilon_{start}$ & 1.0\\
 $\gamma$ & 0.99 \\
 $lr_{cost}$ & $6.03e^{-5}$ \\
 $c_{entropy}$ & $1.66e^{-2}$ \\  
 $lr_{actor}$ & $1e^{-3}$\\
 $lr_{reward}$ & $3e^{-3}$ \\
 $lr_{v}$ & $1e^{-3}$ \\
 $lr_{vmodel}$ & $1e^{-3}$ \\
 $lr_{vrewarder}$ & $1e^{-3}$ \\
 $c_{reg}$ & $1.0$  \\
 $c_{costreg}$ & $1e^{-2}$ \\
 $c_{costreg2}$ & $1e^{-4}$ \\
 \hline
\end{tabular}
\end{table}

\begin{table}[h!]
\centering
\caption{IQ-Flow hyperparameters in Escape Room}
\label{tab:IQER}
\begin{tabular}{ c | c c }
  Parameter & \textbf{ER(5,2)} & \textbf{ER(10,5)} \\ 
  \hline
 $\epsilon_{end}$ & 0.05 & 0.5\\
 $\epsilon_{start}$ & 1.0 & 1.0\\
 $\gamma$ & 0.99 & 0.99\\
 $c_{entropy}$ & $1.66e^{-2}$ & $1.66e^{-2}$\\ 
 $lr_{actor}$ & $9.56e^{-5}$ & $9.56e^{-5}$\\
 $lr_{reward}$ & $1e^{-3}$ & $1e^{-3}$\\
 $lr_{vmodel}$ & $1e^{-3}$ & $1e^{-3}$\\
 $lr_{vrewarder}$ & $1e^{-3}$ & $1e^{-3}$\\
 $c_{costreg}$ & 0.5 & 0.5\\
 $c_{costreg2}$ & 0.5 & 0.5\\
 $lr_{cost}$ & $6.03e^{-5}$ & $6.03e^{-5}$\\
 $h_{1}$ & $64$ & $64$\\  
 $h_{2}$ & $64$ & $64$\\
 $hr_{1}$ & $64$ & $64$\\  
 $hr_{2}$ & $32$ & $32$\\
 \hline
\end{tabular}
\end{table}

\begin{table}[h!]
\centering
\caption{IQ-Flow hyperparameters in 2-Player Cleanup}
\label{tab:IQSSD}
\begin{tabular}{ c | c}
  Parameter & \textbf{}\\ 
  \hline
 $\epsilon_{end}$ & 0.05 \\
 $\epsilon_{start}$ & 1.0 \\
 $\gamma$ & 0.99 \\
 $c_{entropy}$ & $8.1439e^{-2}$\\
 $lr_{actor}$ & $1.959e^{-4}$\\
 $lr_{reward}$ & $4.323e^{-6}$\\
 $lr_{v}$ & $7.445e^{-5}$\\
 $lr_{vmodel}$ & $4.0089e^{-5}$\\
 $lr_{vrewarder}$ & $1e^{-3}$\\
 $c_{costreg}$ & $0.25$ \\
 $c_{costreg2}$ & $1.4924e^{-4}$ \\
 $\tau$ & 0.01 \\
 $h_{1}$ & $64$\\  
 $h_{2}$ & $64$\\
 $filters$ & $6$\\
 $kernel$ & $[3, 3]$\\
 \hline
\end{tabular}
\end{table}

\section{Additional ablation experiments}\label{ablationextra}

\begin{figure}[H]
  \centering
  \includegraphics[width=\linewidth]{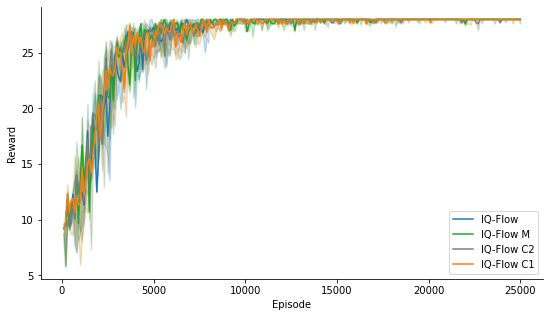}
  \caption{ER(5,2)}
  \label{fig:sub1}
\end{figure}

\begin{figure}[H]
  \centering
  \includegraphics[width=\linewidth]{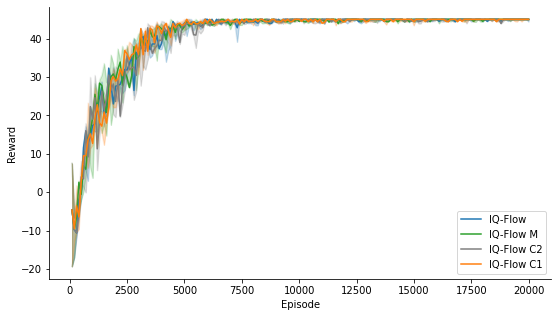}
  \caption{ER(10,5)}
  \label{fig:sub2}
\end{figure}

\section{Iterated Matrix Game IQ-FLow payoff results}\label{ipdfull}

\begin{figure}[H]
\centering
    \includegraphics[width=.8\linewidth]{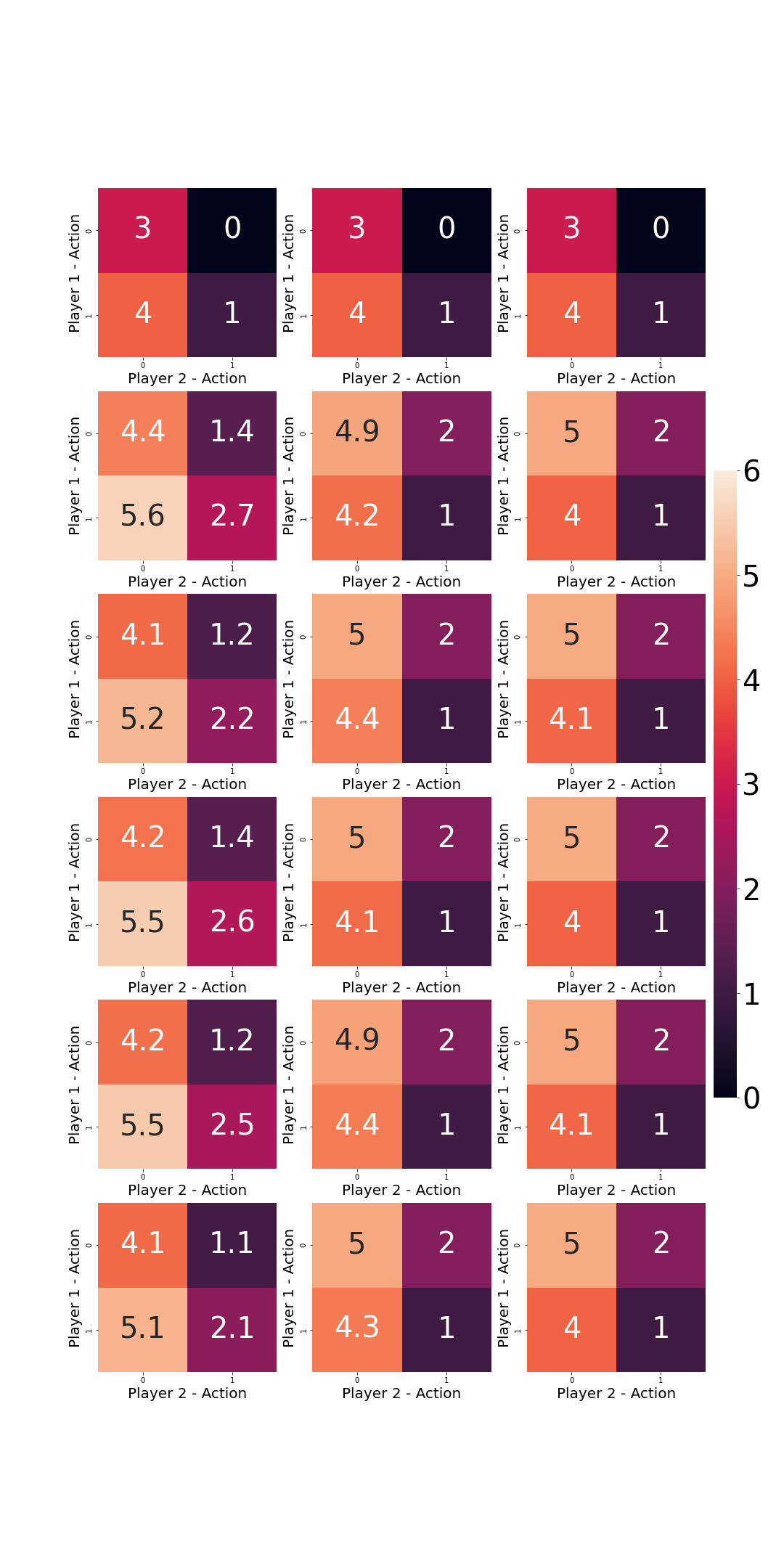}
\caption{IPD Player 1 Payoff matrices}
\label{p1ipdpayofffull}
\end{figure}

\begin{figure}[H]
\centering
    \includegraphics[width=\linewidth]{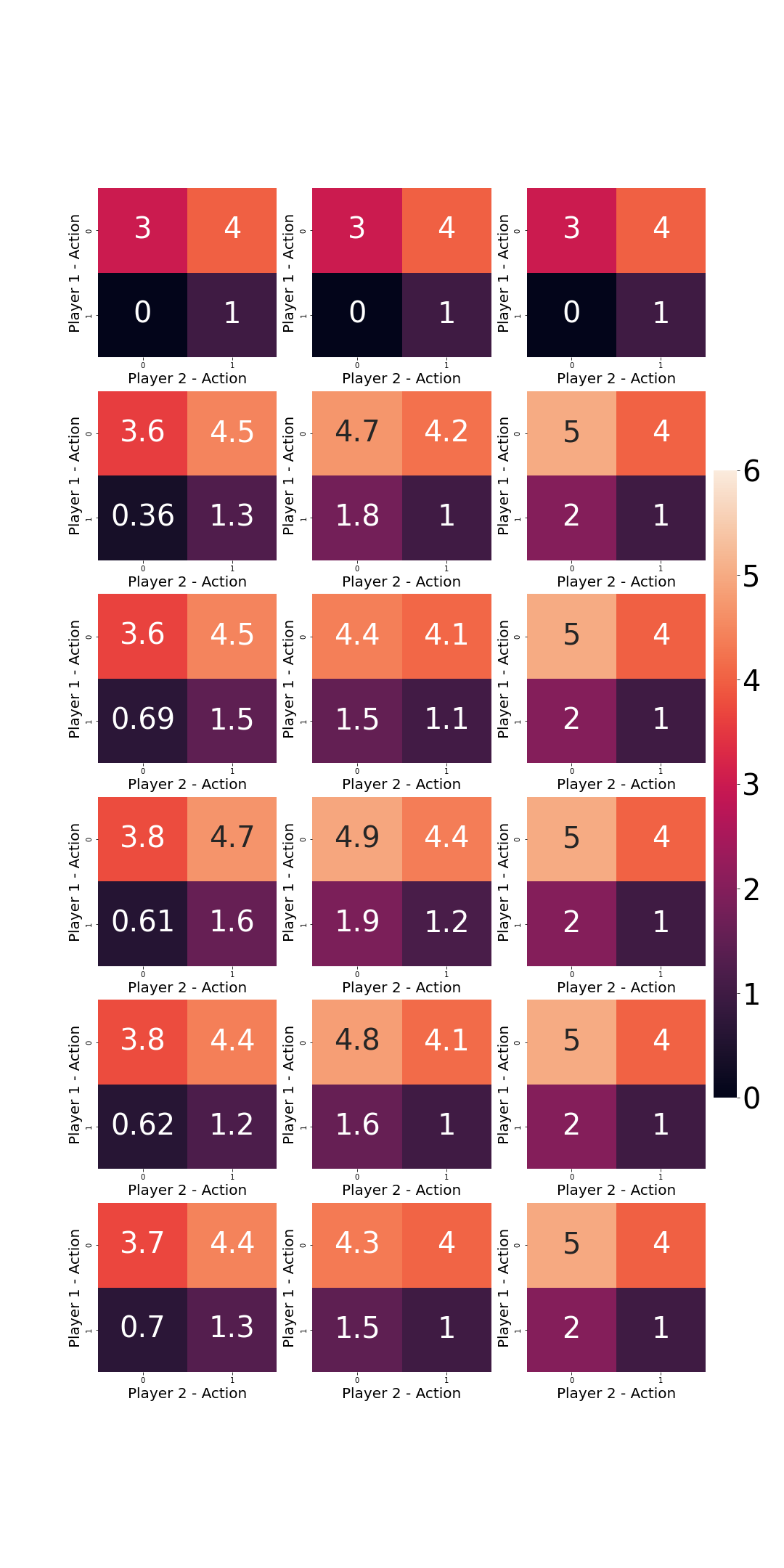}
\caption{IPD Player 2 Payoff matrices}
\label{p2ipdpayofffull}
\end{figure}

\begin{figure}[H]
\centering
    \includegraphics[width=\linewidth]{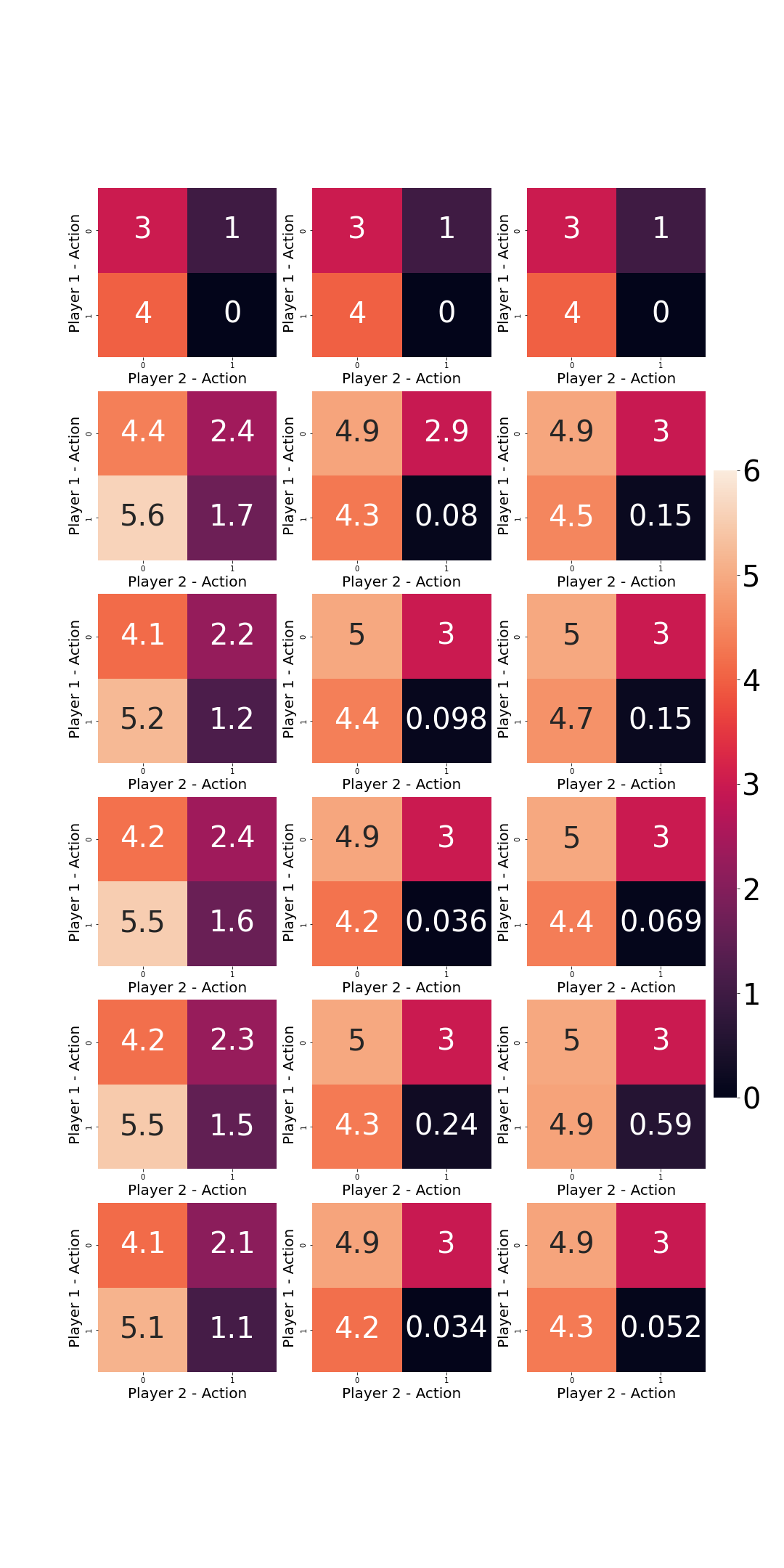}
\caption{Chicken Game Player 1 Payoff matrices}
\label{p1chickpayofffull}
\end{figure}

\begin{figure}[H]
\centering
    \includegraphics[width=\linewidth]{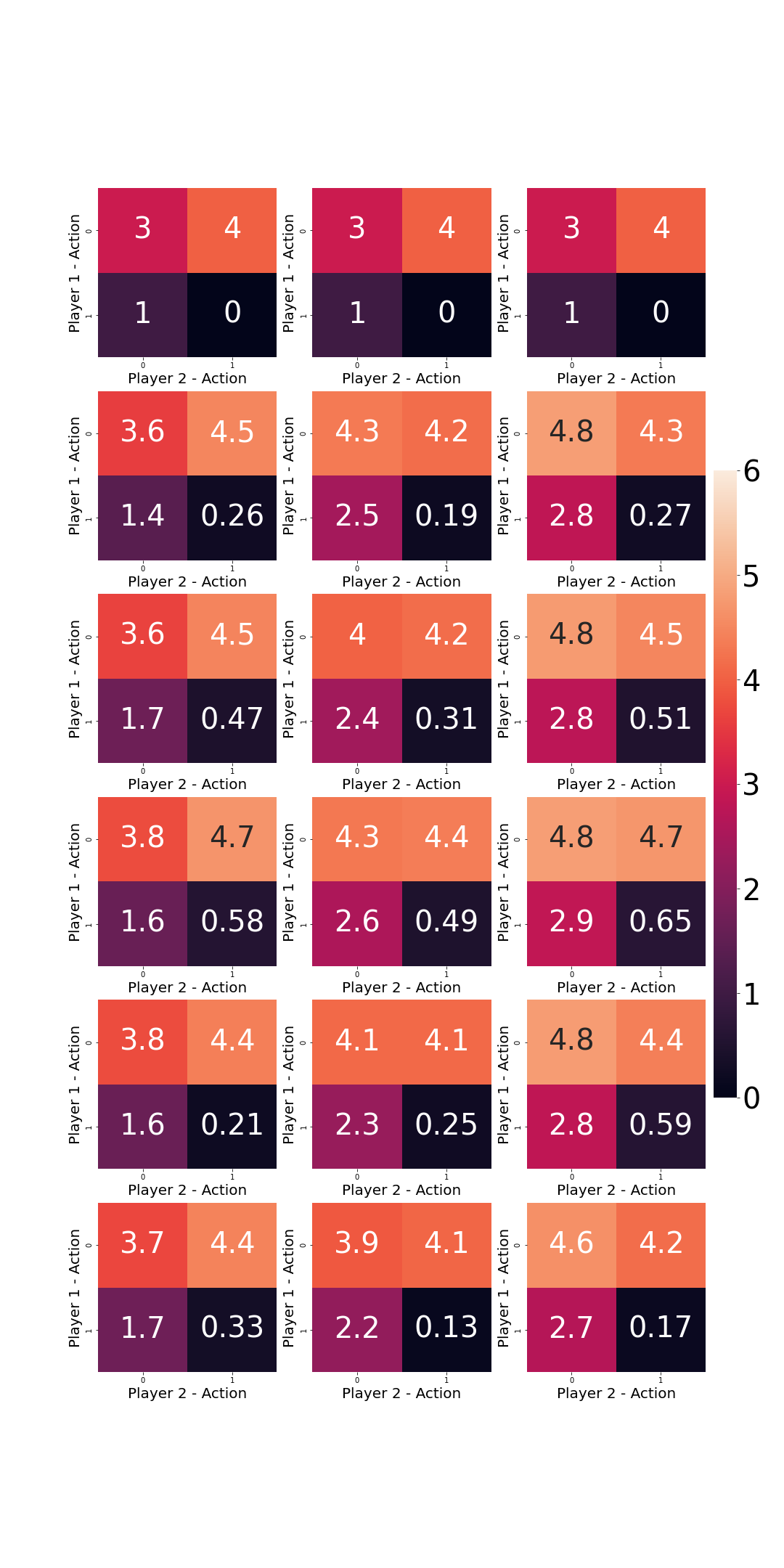}
\caption{Chicken Game Player 2 Payoff matrices}
\label{p2chickpayofffull}
\end{figure}

\begin{figure}[H]
\centering
    \includegraphics[width=\linewidth]{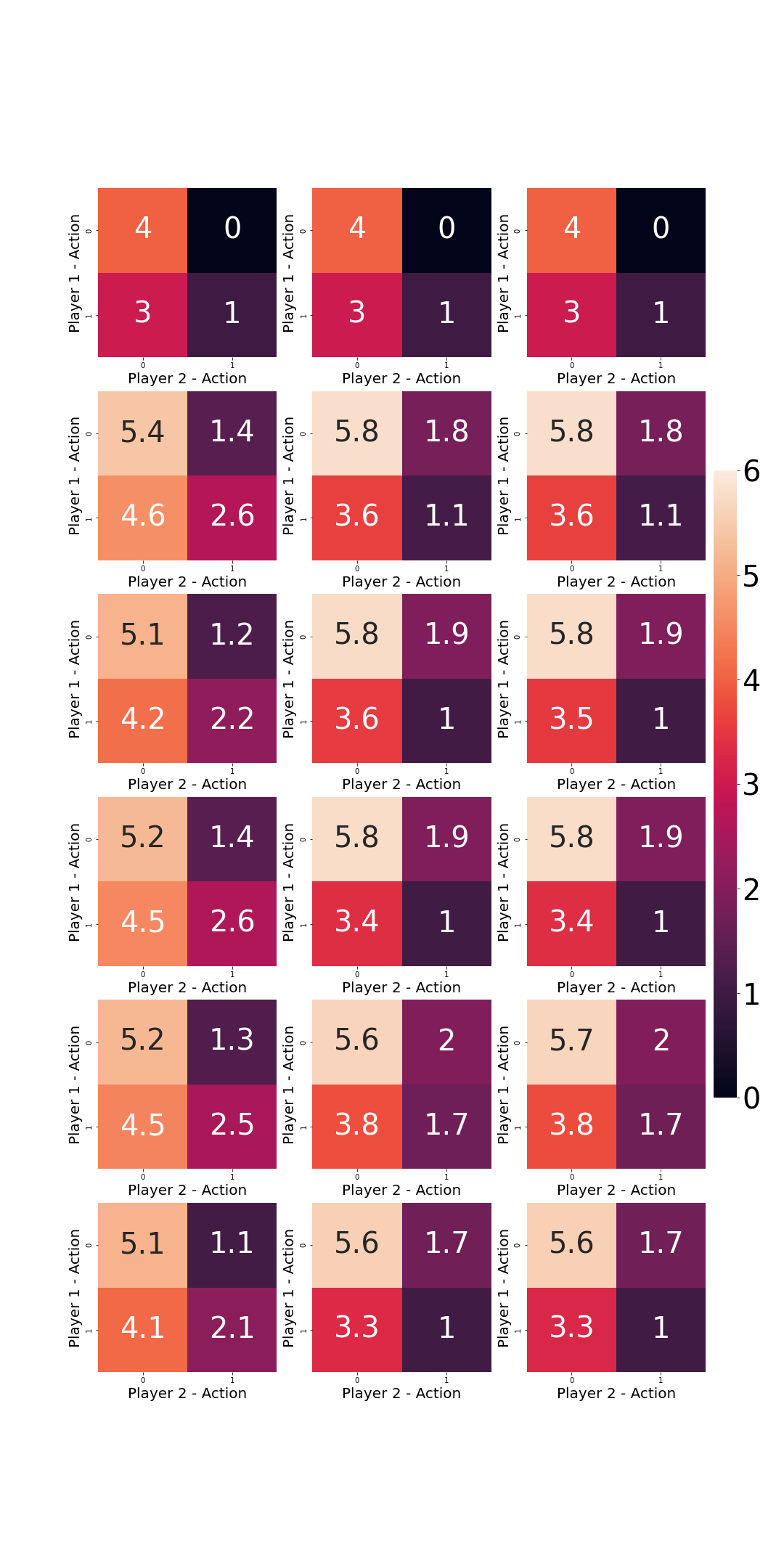}
\caption{Stag Hunt Player 1 Payoff matrices}
\label{p1stagpayofffull}
\end{figure}

\begin{figure}[H]
\centering
    \includegraphics[width=\linewidth]{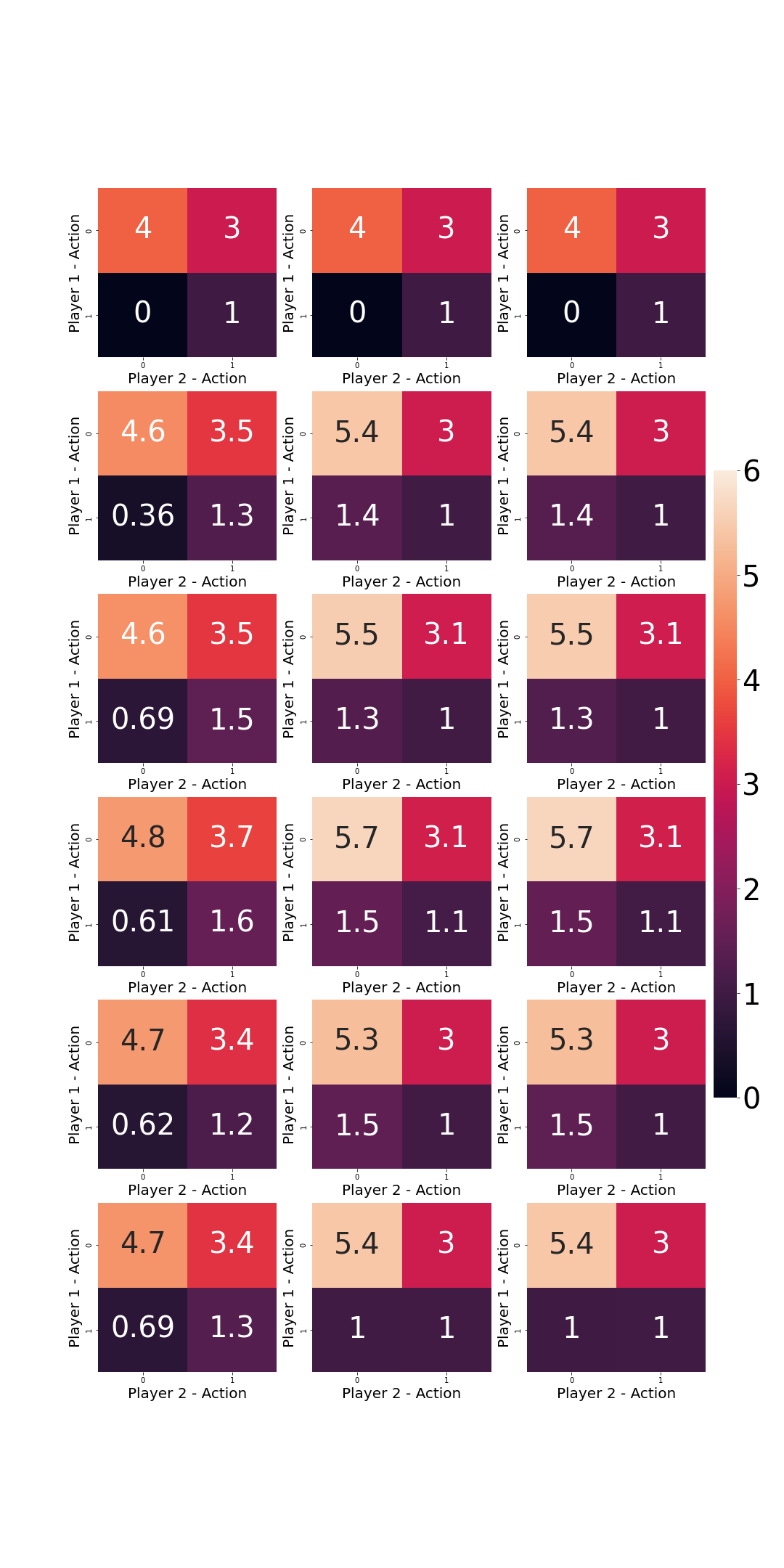}
\caption{Stag Hunt Player 2 Payoff matrices}
\label{p2stagpayofffull}
\end{figure}

\balance

\section{Iterated Matrix Game IQ-FLow Q-Value results}\label{ipdfullqtable}

\begin{figure}[H]
\centering
    \includegraphics[width=.5\linewidth]{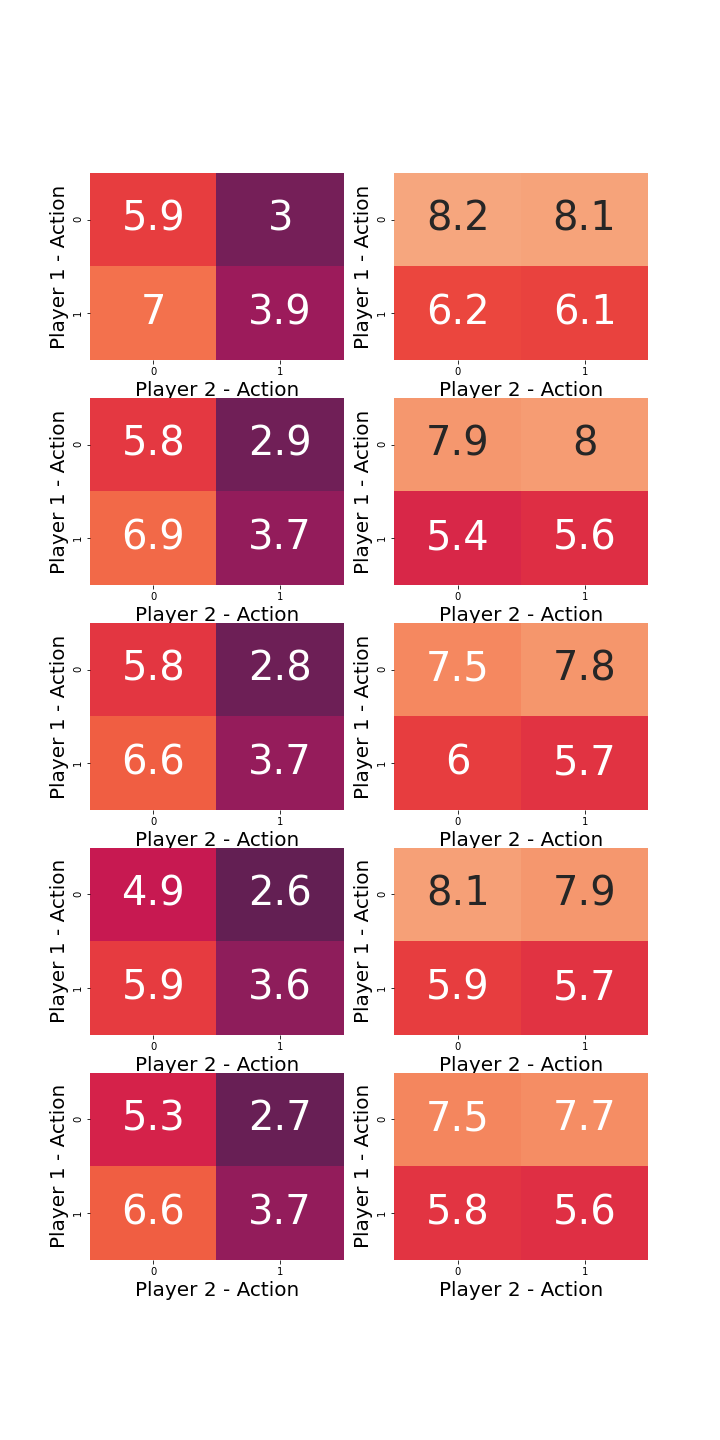}
\caption{IPD Player 1 Q-Values}
\label{p1ipdQpayofffull}
\end{figure}

\begin{figure}[H]
\centering
    \includegraphics[width=.5\linewidth]{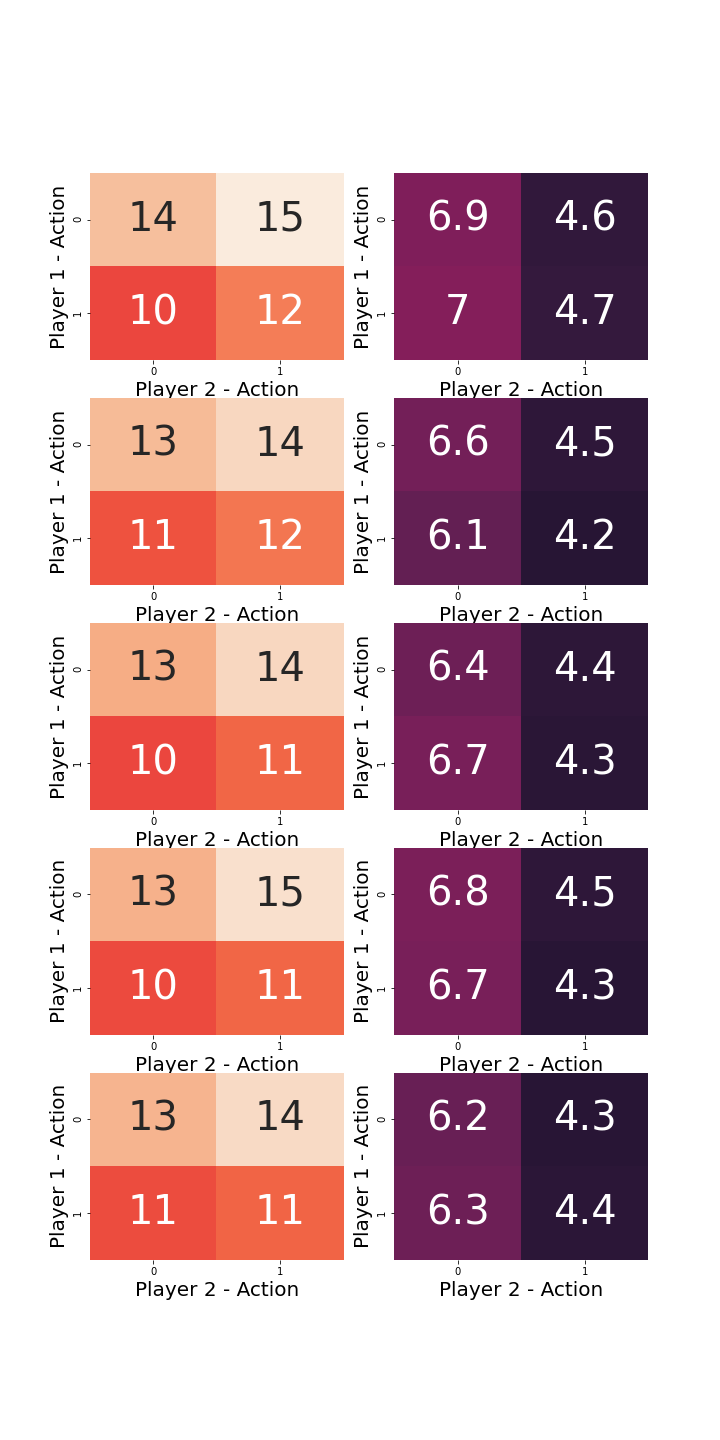}
\caption{IPD Player 2 Q-Values}
\label{p2ipdQpayofffull}
\end{figure}

\begin{figure}[H]
\centering
    \includegraphics[width=.5\linewidth]{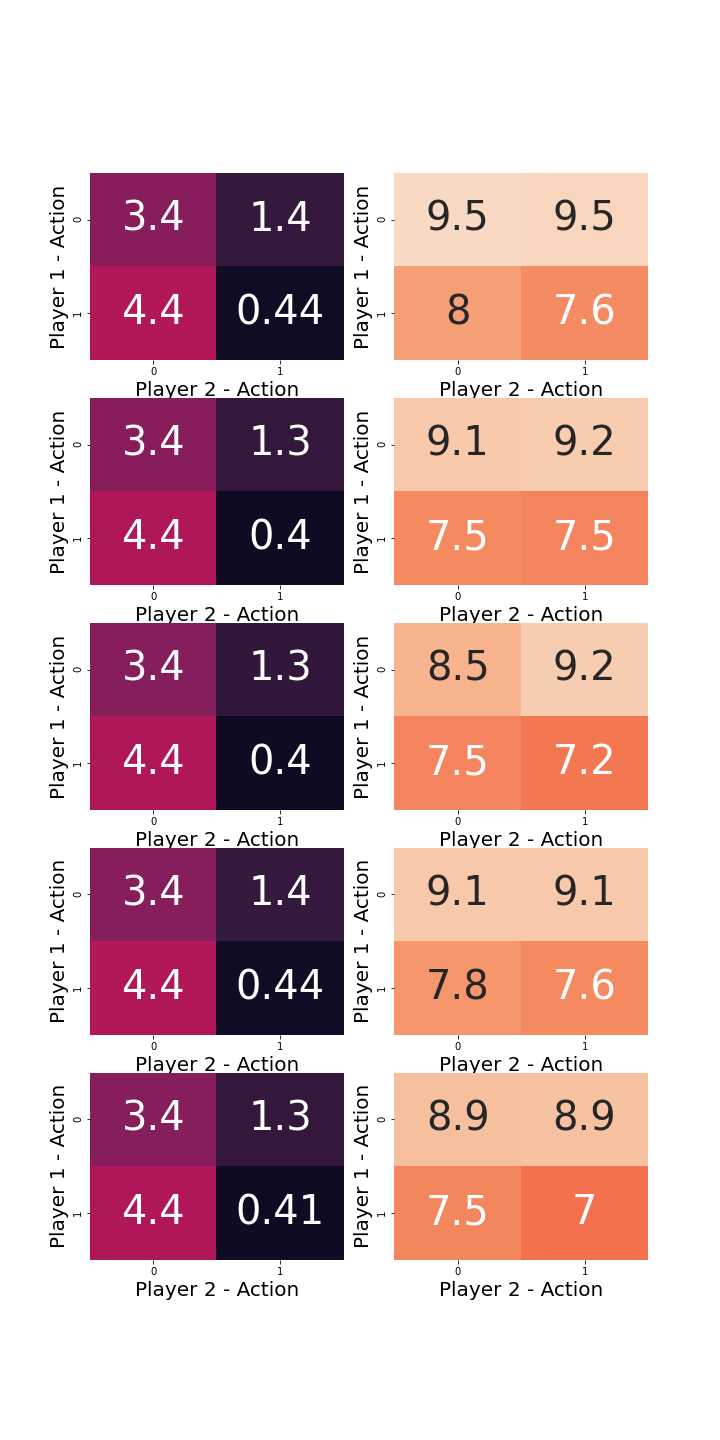}
\caption{Chicken Game Player 1 Q-Values}
\label{p1chickQpayofffull}
\end{figure}

\begin{figure}[H]
\centering
    \includegraphics[width=.5\linewidth]{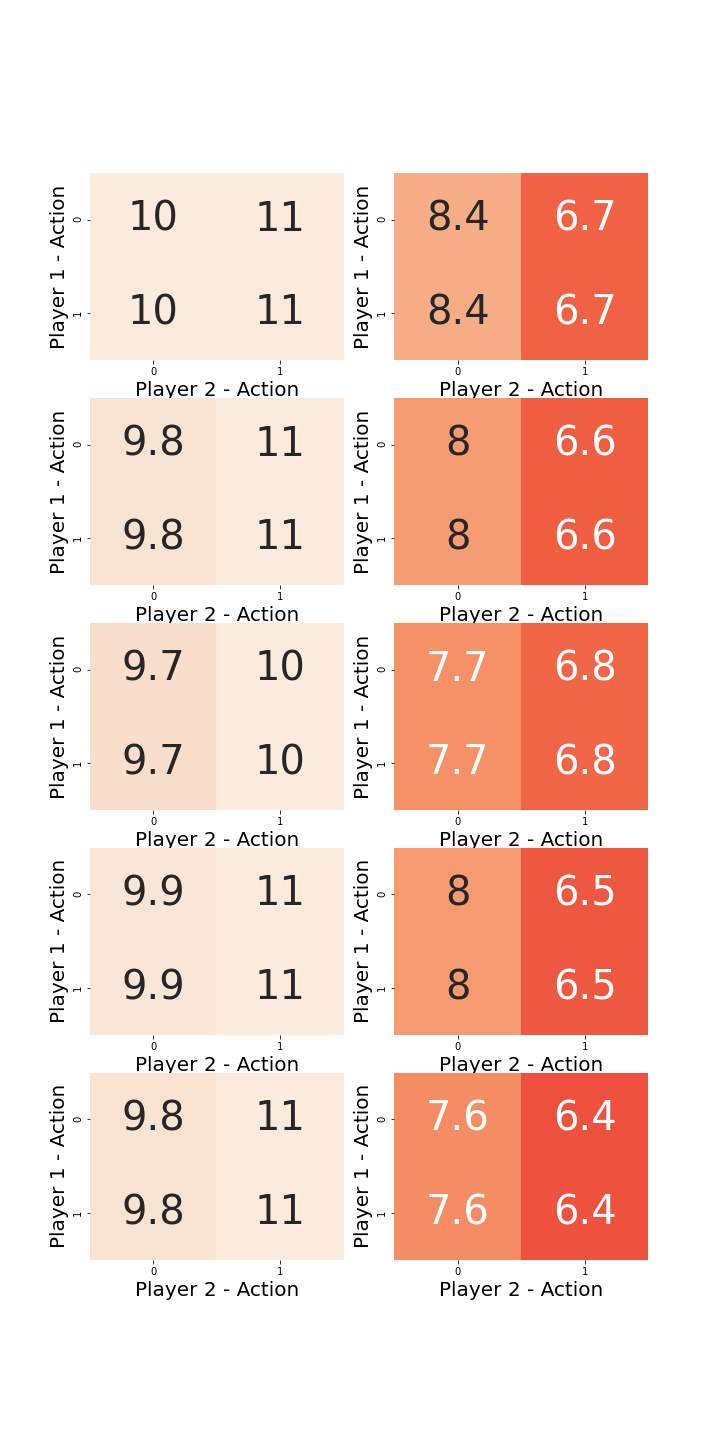}
\caption{Chicken Game Player 2 Q-Values}
\label{p2chickQpayofffull}
\end{figure}

\begin{figure}[H]
\centering
    \includegraphics[width=.5\linewidth]{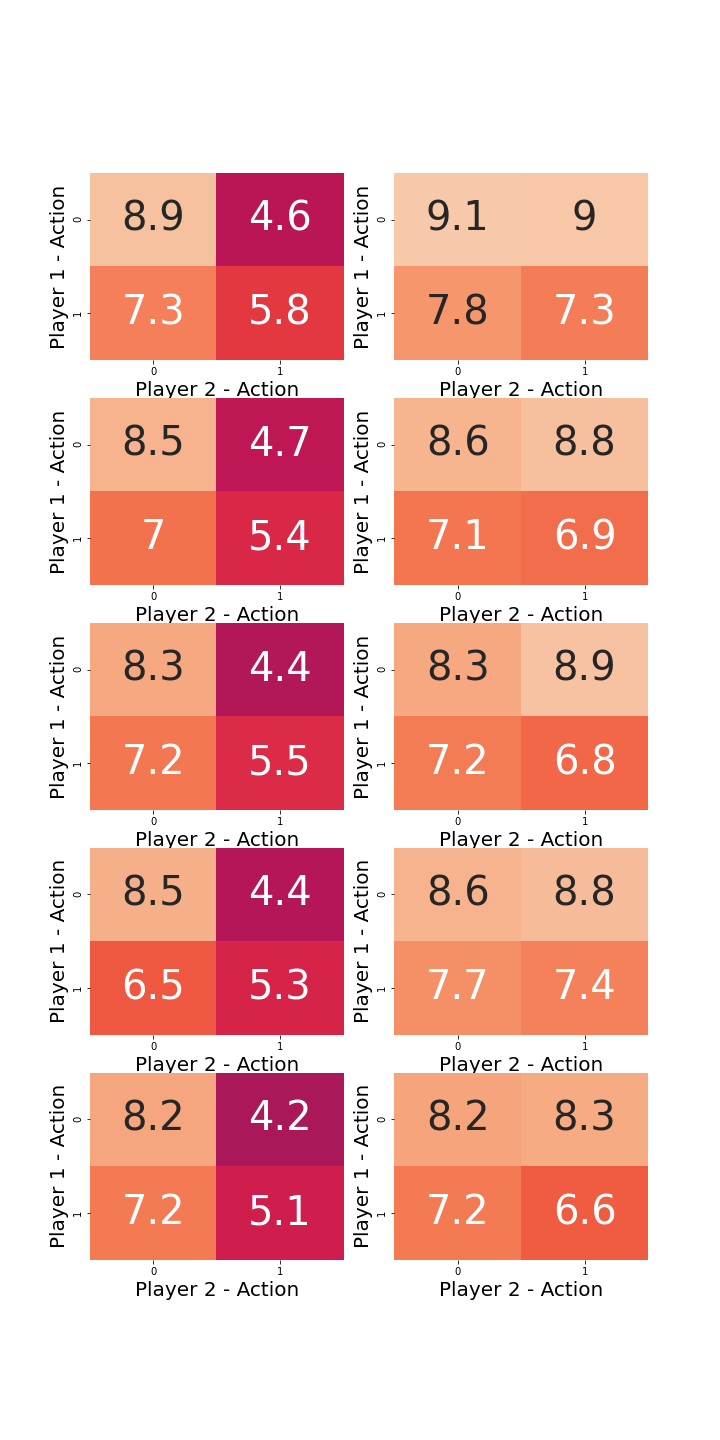}
\caption{Stag Hunt Player 1 Q-Values}
\label{p1stagQpayofffull}
\end{figure}

\begin{figure}[H]
\centering
    \includegraphics[width=.5\linewidth]{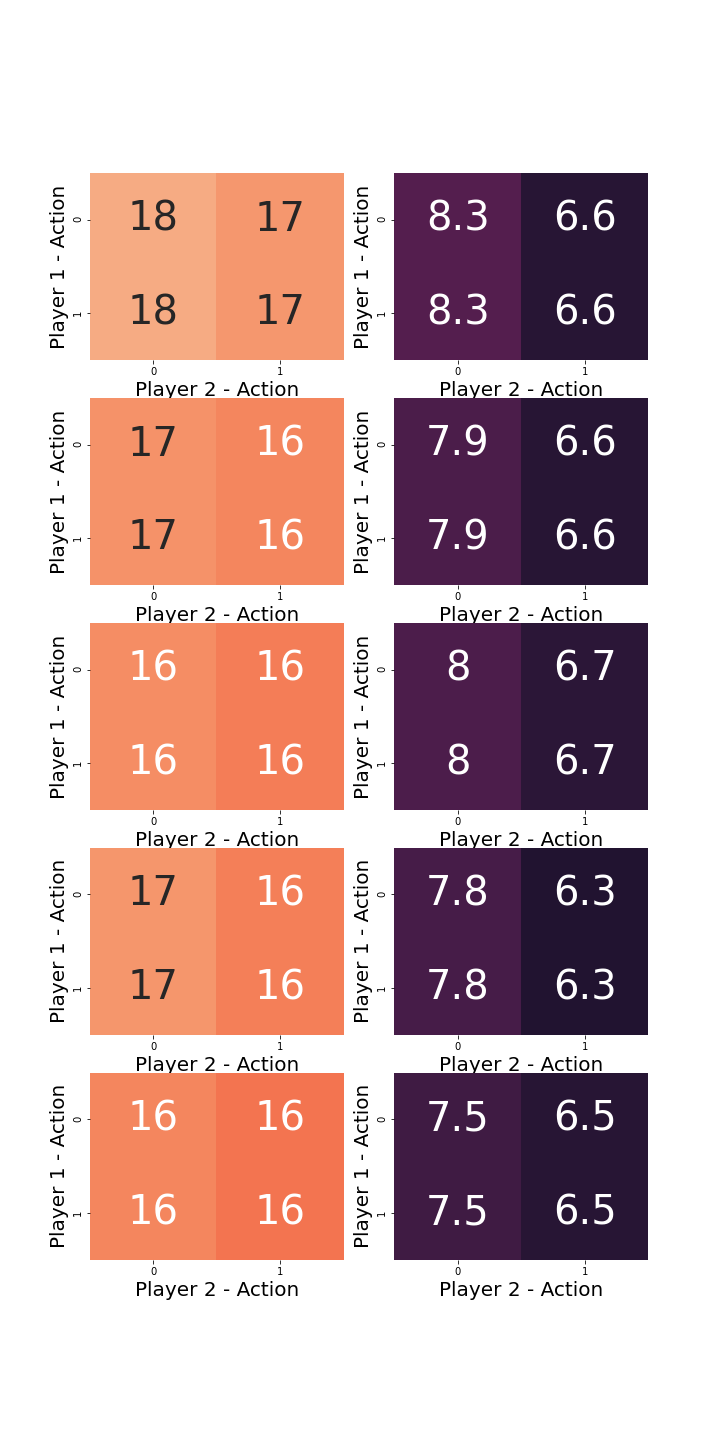}
\caption{Stag Hunt Player 2 Q-Values}
\label{p2stagQpayofffull}
\end{figure}


\balance

\section{Iterated Matrix Games IQ-FLow R - T and S - P Plots for Q-Values}\label{ipdineq}

\begin{figure}[H]
\centering
    \includegraphics[scale=0.22]{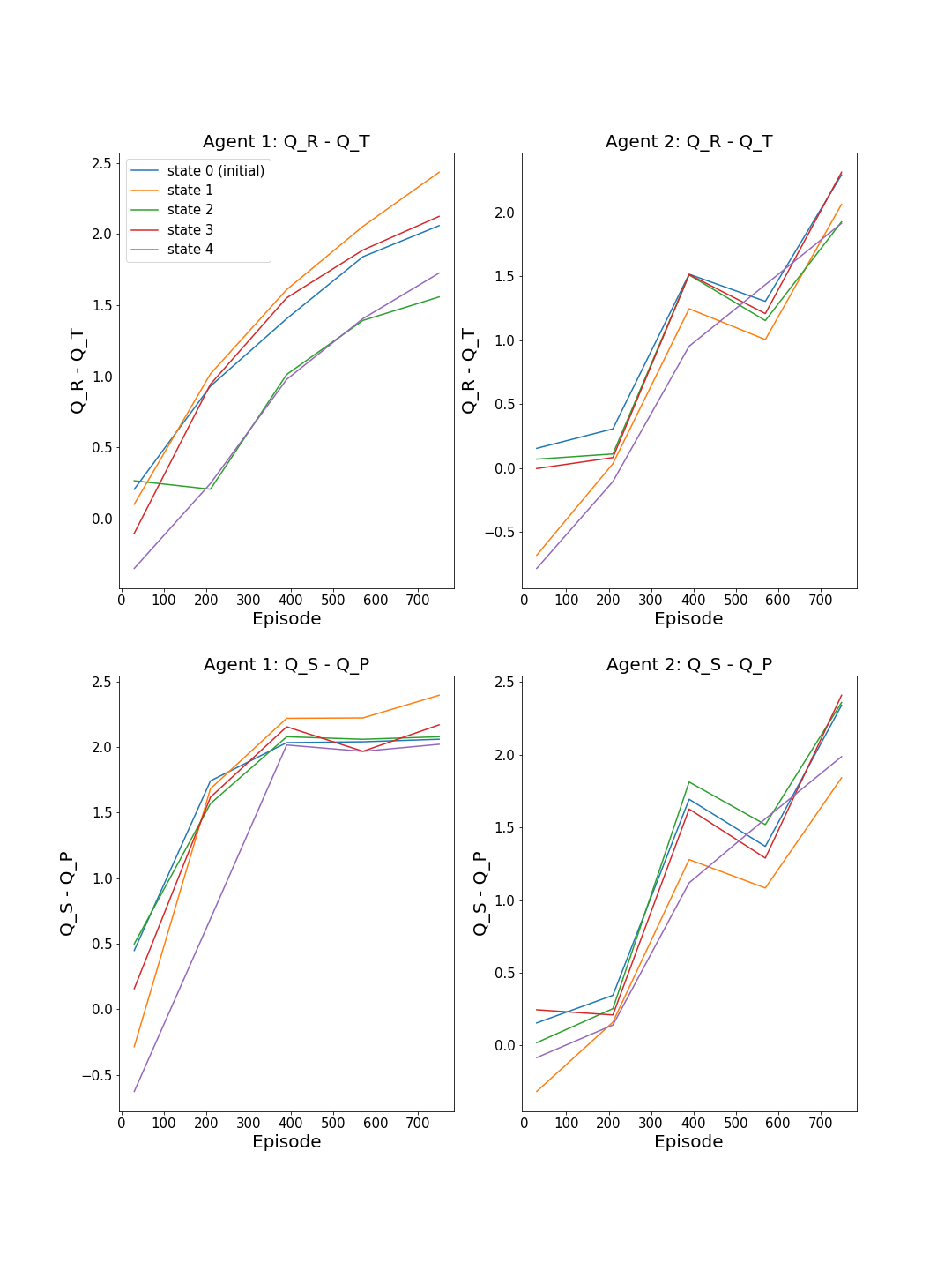}
\caption{IPD $R - T$ and $S - P$ plot for Q-Values}
\label{ipdineqq}
\end{figure}

\balance

\begin{figure}[H]
\centering
    \includegraphics[scale=0.22]{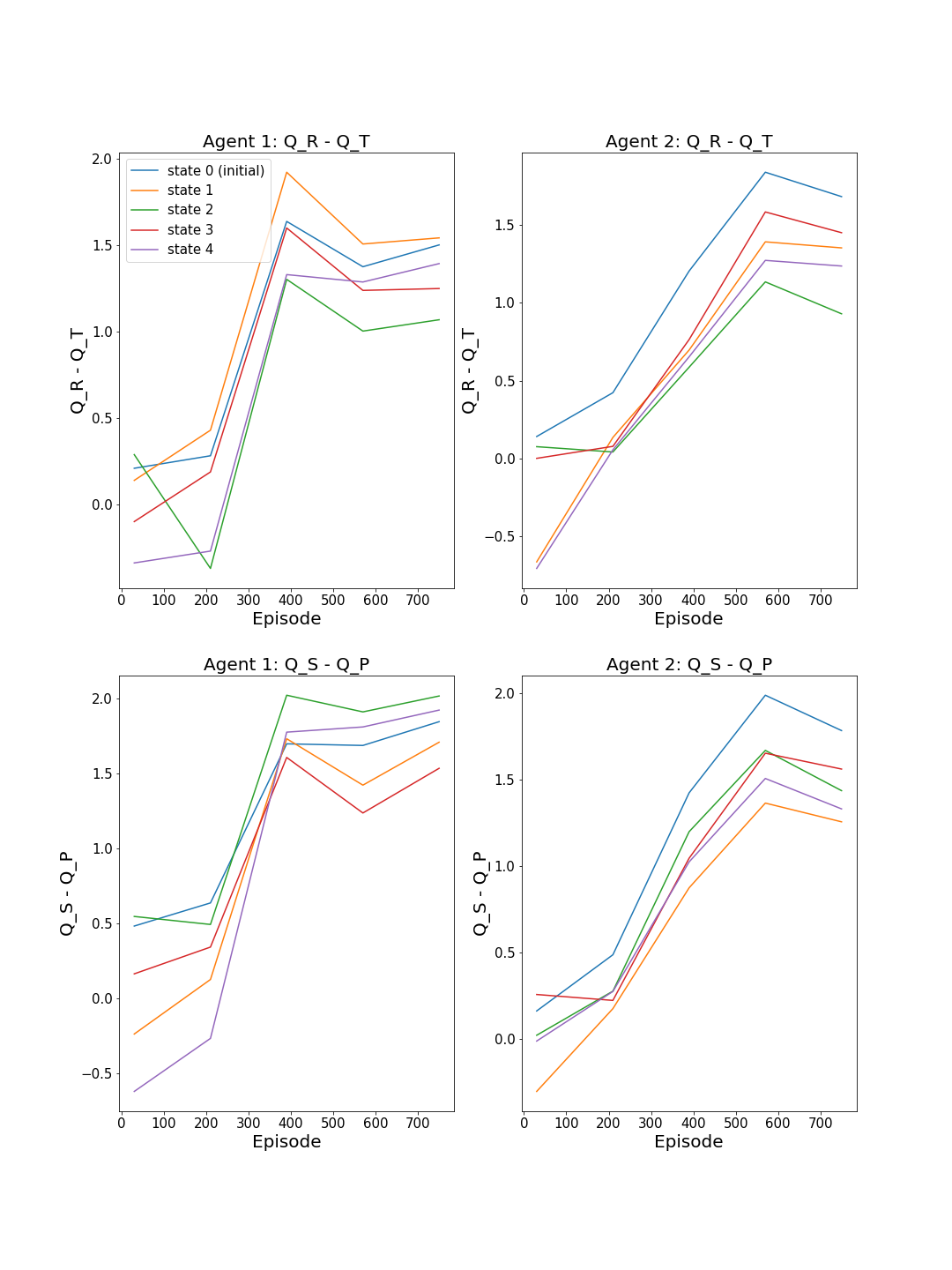}
\caption{Chicken Game $R - T$ and $S - P$ plot for Q-Values}
\label{chickineqq}
\end{figure}

\begin{figure}[H]
\centering
    \includegraphics[scale=0.22]{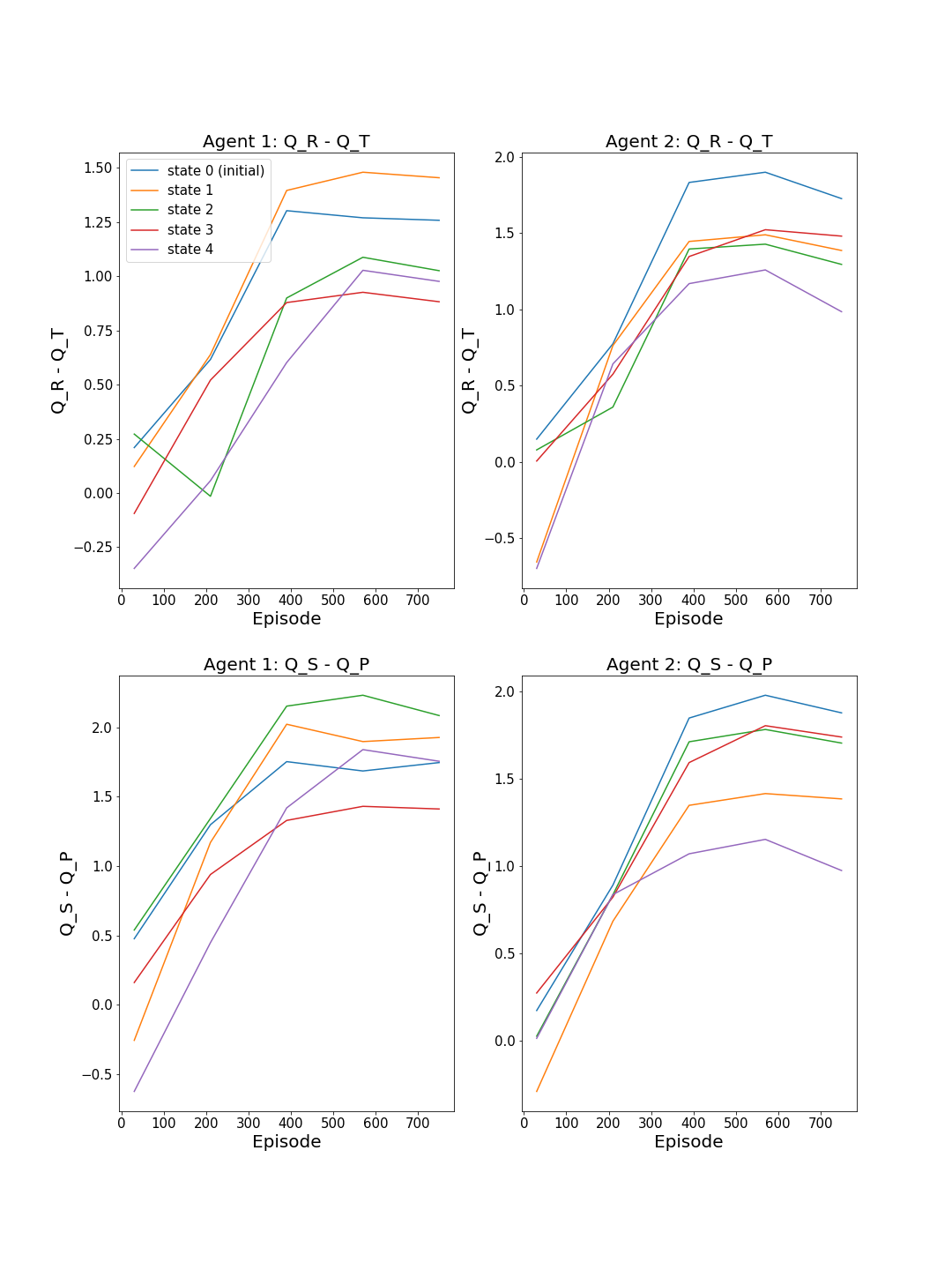}
\caption{Stag Hunt $R - T$ and $S - P$ plot for Q-Values}
\label{stagineqq}
\end{figure}


\end{document}